\documentclass[aps,pre,showpacs,twocolumn]{revtex4}
\usepackage{bm}
\usepackage{graphicx}
\bibstyle{approve.bib}
\usepackage{amssymb}
\usepackage{amsmath}
\usepackage{esint}
\usepackage{epstopdf}
\usepackage{color}
\usepackage{gensymb}
\usepackage{mathtools}
\usepackage{suffix}

\newcommand{\be}{\begin{equation}}
\newcommand{\ee}{\end{equation}}
\newcommand{\bea}{\begin{eqnarray}}
\newcommand{\eea}{\end{eqnarray}}
\newcommand{\lan}{\left\langle}
\newcommand{\ran}{\right\rangle}
\newcommand{\br}{\mathbf{r}}
\newcommand{\ba}{\bm{a}}

\newcommand{\bA}{\mathbf{A}}

\newcommand{\tu}{\tilde{U}}

\newcommand{\bb}{\mathbf{b}}
\newcommand{\pd}{\phi_{\rm D}}

\newcommand{\bu}{\mathbf{u}}

\newcommand{\pn}{P_{\rm net}}

\newcommand{\bk}{\mathbf{k}}

\newcommand{\e}{\varepsilon}

\newcommand{\bd}{\bar{d}}
\newcommand{\pa}{\parallel}

\newcommand{\ce}{_{\rm c}}

\newcommand{\phm}{\phi_0}
\newcommand{\bs}{{\mathbf{S}}}

\newcommand{\bz}{b_{ z}}

\newcommand{\ew}{\varepsilon_{\rm w}}

\newcommand{\SB}[1]{\textcolor{red} {#1}}

\begin{document}

\title{Impact of the inner solute structure on the electrostatic mean-field and strong-coupling regimes of macromolecular interactions}
					
\author{Sahin Buyukdagli\footnote{email:~\texttt{buyukdagli@fen.bilkent.edu.tr}}}
\affiliation{Department of Physics, Bilkent University, Ankara 06800, Turkey}
\date{\today}

\begin{abstract}

The structural diversity of the solute molecules involved in biomolecular processes necessitates the characterization of the forces between charged macromolecules beyond the point-ion description. From the field theoretic partition function of an electrolyte confined between two anionic membranes, we derive a contact-value identity valid for general intramolecular solute structure and electrostatic coupling strength. In the electrostatic mean-field (MF) regime, the inner charge spread of the solute particles is shown to induce the twofold enhancement of the short-range Poisson-Boltzmann (PB)-level membrane repulsion, and a longer-range depletion attraction. Our contact theorem indicates that  the twofold repulsion enhancement by solute size is equally present in the opposite strong-coupling (SC) regime of linear and spherical solute molecules. Upon the inclusion of the dielectric contrast between the electrolyte and the interacting membranes, the emerging polarization forces substantially amplify the solute specificity of the macromolecular interactions. Namely, the finite size of the solute particles composed of similar terminal charges such as putrescine molecules weaken the intermembrane repulsion. However, the extended structure of the solute molecules carrying opposite elementary charges such as ionized atoms and zwitterionic molecules enhance the membrane repulsion by several factors.  We also show that these polarization forces can extend the range of the solute structure effects up to intermembrane distances exceeding the solute size by an order of magnitude. This radical alteration of the intermembrane interactions by the salt structure identifies the solute specificity as a key ingredient of the thermodynamic stability in colloidal systems.

\end{abstract}

\date{\today}
\maketitle

\section{Introduction}

The internal structure of the microscopic particles invisible to the naked eye is often discernible by the macroscopic behavior mediated by them~\cite{Isr}. The salt specificity of protein stability~\cite{Hof1} and the surface tension of electrolyte solutions~\cite{Levin1} obeying the Hofmeister series are notorious examples of this upward causation~\cite{SurTenHof}.  The large-scale macromolecular aggregation triggered by depletion interactions is another illustration of the macroscopic behavior affected by the intramolecular solute structure. The depletion forces induced by the excluded volume of the solute particles are frequently used in industrial applications requiring phase separation~\cite{Rev3,Rev1}, such as paint making~\cite{Rev2} and water purification by flocculation~\cite{pur}.

The first formulation of the depletion interactions has been introduced by Asakura and Oosawa in 1954~\cite{Oosawa1}. The Asakura-Oosawa (A-O) theory predicts that if two uncharged macromolecules suspended in a solution of smaller particles approach each other by a distance shorter than the size of these particles, the sterically induced particle exclusion from the intermolecular region leads to an inward osmotic pressure gradient driving the macromolecules towards each other. Subsequent works extended the A-O model to interacting charged macromolecules immersed in a solution of spherical particles~\cite{deplCh}, formulated the problem within the framework of the density functional theory~\cite{deplDFT}, and investigated the depletion interactions induced by polymeric depletants~\cite{deplPOLY1,deplPOLY2}. Depletion attraction has been equally studied experimentally via atomic force microscope and light scattering techniques~\cite{exp0,exp1,exp2,exp3}. The history of the discovery and the current applications of this entropic mechanism are elaborated in Refs.~\cite{Rev1,Rev2,Rev3}.

Despite originating from a fundamentally distinct physical mechanism, the outcome of the electromagnetic Casimir interactions between two plates located in a vacuum environment bears some qualitative similarities with the aforementioned depletion attractions. In a parallel plate configuration, the exclusion of the long-wavelength modes from the interplate zone leads to the mutual attraction of the plates even in the absence of any excess charge on their surface~\cite{casth1,casth2,casex1,casex2}. The thermal van der Waals (vdW) interactions occurring between solvated macromolecules correspond precisely to the classical counterpart of the Casimir interactions ~\cite{Lifshitz,Derja,Tabor1,Tabor2,Ninh,Hough}.  The interplay between the short-ranged vdW attraction and the longer-ranged double layer repulsion~\cite{Levine1,Levine2} is the key ingredient of the Derjaguin-Landau-Verywey-Overbeek (DLVO) theory describing the electrostatic equilibrium of macromolecules in salty solutions~\cite{DLVO}. 

The primitive model at the basis of the DLVO formalism is limited by its implicit solvent framework and the underlying point-ion assumption. Explicit solvent structure and ionic polarizability have been introduced into the primitive model by point-dipole approaches~\cite{dunyuk,orland1,podgornik,podgornik2}. In order to relax the point-dipole approximation, we have previously developed a non-local electrostatic theory of finite-size solvent molecules and polarizable ions in contact with single interfaces~\cite{PRE1}. Within this non-local PB (NLPB) framework, we have recently characterized the effect of explicit solvent on macromolecular interactions~\cite{PRE2022}. Additional works focused on the electrostatics of rigid particles to explain the ordering transition in bulk liquids~\cite{Lue1}, and the bridging attraction between like-charged macromolecules~\cite{bohinc1,bohinc2,bohinc3,bohinc4,pincus}.

In this article, we characterize the effect of the intramolecular solute structure on like-charge intermembrane interactions. The novelties of our formalism with respect to earlier works  are (i) the derivation of an exact contact theorem for general solute structure, (ii) the incorporation of the inner solute charges without any multipolar expansion, (iii) the evaluation of the SC-level intermembrane pressure mediated by spherical solute molecules, and (iv) the characterization of the solute structure effects on the coupling of low permittivity membranes governed by dielectric polarization forces.

Our manuscript is organized as follows. In Section~\ref{model}, we compute the field theoretic partition function of a structured electrolyte confined between adjacent anionic membranes. From the liquid partition function, we derive a generalized contact-value identity valid for arbitrary solute charge structure and electrostatic coupling strength. In the present article, the corresponding formalism is considered for linear and spherical solute molecules illustrated in Figs.~\ref{fig1}(b)-(c). In Section~\ref{res}, we characterize the effect of the linear solute structure on the electrostatic MF-regime of intermembrane interactions. We show that the inner charge spread of the solute molecules gives rise to the twofold enhancement of the short-range PB-level repulsion, which is followed by a longer-range depletion attraction between the like-charged membranes. In Sec.~\ref{sc}, we find that the twofold repulsion enhancement is equally present in the opposite electrostatic regime of SC interactions for both linear and spherical solute particles. The corresponding pressure profiles are schematically displayed in Figs.~\ref{fig1}(d)-(e). Finally, we probe the solute structure effects in the presence of the dielectric contrast between the liquid and the typically low permittivity membranes. We show that the emerging polarization forces enhance significantly the impact and range of the solute specificity on the intermembrane coupling. In Conclusions, we summarize our findings, and discuss the limitations and potential extensions of our formalism.

\section{Field Theory of Structured Liquids}
\label{model}

\subsection{Field-theoretic partition function}
\label{th}

\begin{figure}
\includegraphics[width=1\linewidth]{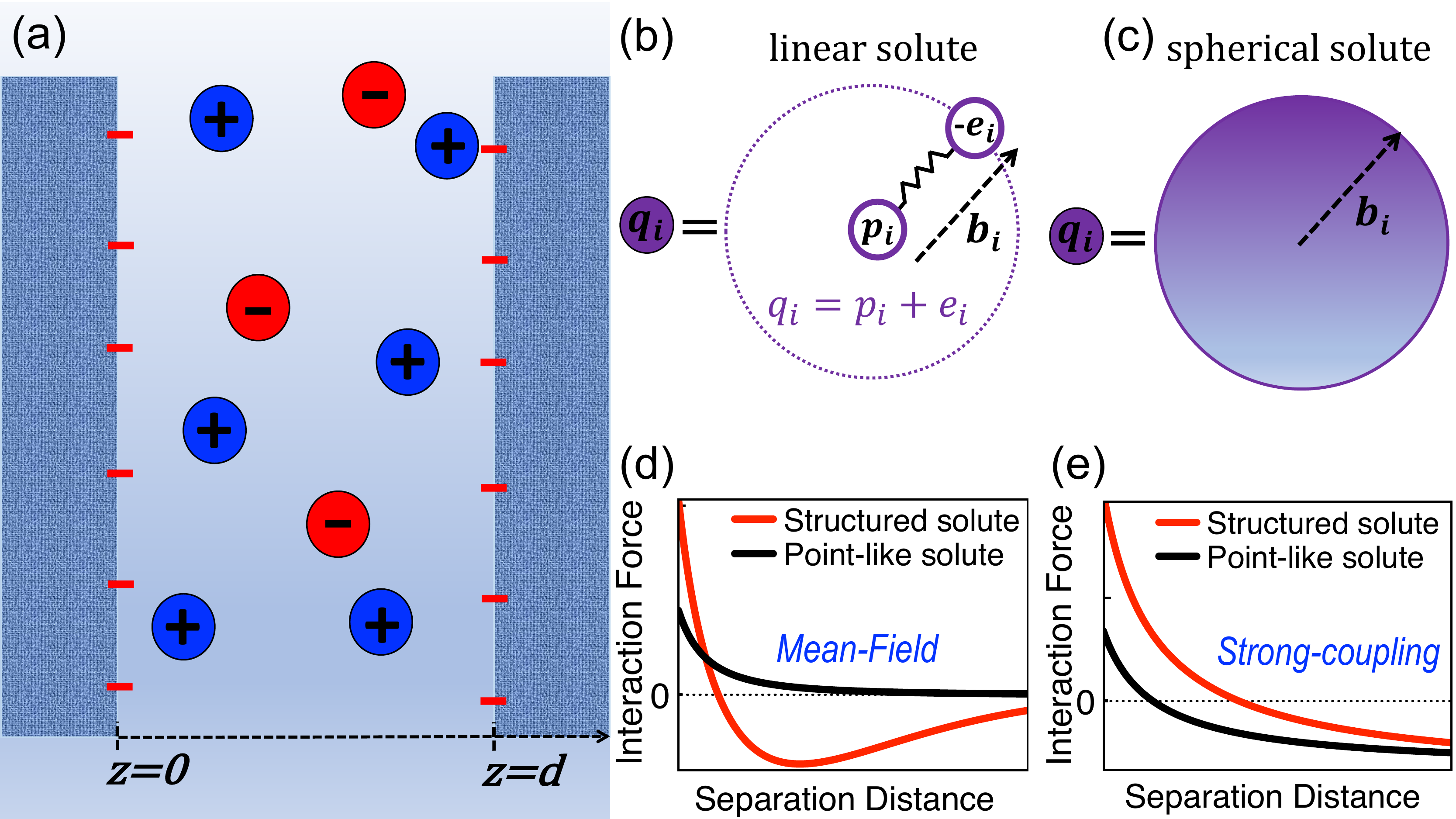}
\caption{(Color online) (a) Depiction of the anionic membranes immersed in an electrolyte composed of solute molecules with valencies $q_i$. (b)-(c) Inner structure of the salt charges. The linear solute molecules are composed of two terminal charges $p_i$ and $e_i$, separated by the fluctuating distance $\bb_i$ of average magnitude $a_i$ and variance $\alpha_i$. The spherical solute molecules with fluctuating radius $b_i$ carry an isotropic surface charge. (d)-(e) Schematic representation of the intermembrane force with point-like (black) and linear solute charges (red)\cite{rem1}.}
\label{fig1}
\end{figure}

The geometry of the inhomogeneous liquid is displayed in Fig.~\ref{fig1}(a). The system is composed of two anionic membranes with separation distance $d$ and large lateral surfaces $S\gg d^2$. The fixed membrane charge distribution of surface density $-\sigma_{\rm m}<0$ is given by
\be\label{e10}
\sigma(\br)=\sigma(z)=-\sigma_{\rm m}\left[\delta(z)+\delta(d-z)\right],
\ee
with the Dirac delta function $\delta(x)$~\cite{math}. These charged membranes are immersed in an electrolyte solution. The electrolyte is composed of an implicit solvent modeled as a dielectric continuum of relative permittivity $\ew=78$, and $s$ species of solute molecules. Unless stated otherwise, we will neglect the spatial variations of the dielectric permittivity and take $\e(\br)=\e_0\ew$ for the entire system, where $\e_0$ stands for the vacuum permittivity. The liquid confined between the membrane walls at $z=0$ and $z=d$ is in chemical equilibrium with a bulk reservoir of the same solution. In the liquid, the solute molecules of the species $i$ have fugacity $\Lambda_i$ and bulk concentration $\rho_{i{\rm b}}$. The system is at the ambient temperature $T=300$ K. 

The grand-canonical partition function of the structured liquid reads
\be\label{e1}
Z_{\rm G}=\prod_{i=1}^s\sum_{N_i\geq0}\frac{\Lambda_i^{N_i}}{c_iN_i!}\prod_{j=1}^{N_i}\int\mathrm{d}^3\br_{ij}\mathrm{d}^3\bb_{ij}e^{-\beta E-\beta U_i(\bb_{ij})},
\ee
where $\br_{ij}$ stands for the center-of-mass (C.M.) position of the solute molecule $j$ of the species $i$, and $\bb_{ij}$ is the variable characterizing its fluctuating size. The corresponding intramolecular interactions are incorporated via the potential energy $U_i(\bb_{ij})$. The form of this function and the normalization factor $c_i$ will be specified below.

In Eq.~(\ref{e1}), the coupling energy between the charges has been defined as
\bea
\label{e2}
\beta E&=&\frac{1}{2}\int{\rm d}^3\br{\rm d}^3\br'\left[\hat\rho\ce+\sigma\right]_\br v\ce(\br,\br')\left[\hat\rho\ce+\sigma\right]_{\br'}\\
&&+\sum_{i=1}^s\sum_{j=1}^{N_i}\left[w_i(\br_{ij},\bb_{ij})-\epsilon_i(\bb_{ij})\right]\nonumber\\
&&+\int\mathrm{d}^3\br f\ce(\br)\hat\rho\ce(\br)\nonumber
\eea
with the inverse thermal energy $\beta=1/(k_{\rm B}T)$ including the Boltzmann constant $k_{\rm B}$. The first term on the r.h.s. of Eq.~(\ref{e2}) corresponds to pairwise Coulombic interactions between the fixed membrane charges of density $\sigma(\br)$, and the solute molecules of general charge structure function $\hat n_i(\br;\br_{ij},\bb_{ij})$ and density operator
\be\label{solc}
\hat\rho\ce(\br)=\sum_{i=1}^s\sum_{j=1}^{N_i}\hat n_i(\br;\br_{ij},\bb_{ij}).
\ee
The Green's function setting these electrostatic interactions is defined in terms of its inverse as $v\ce^{-1}(\br,\br')=-(k_{\rm B}T/e^2)\nabla\cdot\e(\br)\nabla\delta^3(\br-\br')$, where $e$ is the electron charge. From the identity $\int_{\br''} v^{-1}\ce(\br,\br'')v\ce(\br'',\br')=\delta^3(\br-\br')$, the corresponding kernel equation follows as
\be
\label{greq}
\frac{k_{\rm B}T}{e^2}\nabla\cdot\e(\br)\nabla v\ce(\br,\br')=-\delta^3(\br-\br').
\ee
In the case of dielectrically uniform systems, Eq.~(\ref{greq}) is solved by the bulk Coulomb potential $v_{\rm c,b}(r)=\ell_{\rm B}/r$, with the Bjerrum length $\ell_{\rm B}=e^2/(4\pi\e_0\ew k_{\rm B}T)\approx 7$ {\AA}.

The second term of Eq.~(\ref{e2}) incorporates the steric potential $w_i(\br,\bb)$ acting on the solute molecules. This potential will allow to derive the density of the solute molecules and to impose their confinement to the intermembrane region. The corresponding term also substracts from Eq.~(\ref{e2}) the self-energy of the solute charges,
\be\label{se}
\epsilon_i=\frac{1}{2}\int\mathrm{d}^3\br_1\mathrm{d}^3\br_2\hat n_i(\br_1;\br,\bb)v_{\rm c,b}(\br_1-\br_2)\hat n_i(\br_2;\br,\bb).
\ee
Finally, the third term of Eq.~(\ref{e2}) introduces the generating function $f\ce(\br)$ that will be used to compute the average value of the solute charge density $\hat\rho\ce(\br)$.

Applying to Eq.~(\ref{e1}) a standard Hubbard-Stratonovich transformation, the grand-canonical partition function takes the form of a functional integral over the fluctuating electrostatic potential $\phi(\br)$, 
\be\label{par}
Z_{\rm G}=\int\frac{\mathcal{D}\phi}{\sqrt{{\rm det}\left(v\ce\right)}}\;e^{-\beta H[\phi]}, 
\ee
with the Hamiltonian functional
\bea
\label{e5}
\hspace{0mm}\beta H[\phi]&=&\frac{k_{\rm B}T}{2e^2}\int\mathrm{d}^3\br\;\e(\br)\left[\nabla\phi(\br)\right]^2-i\int\mathrm{d}^3\br\sigma(\br)\phi(\br)\nonumber\\
&&-\sum_{i=1}^s\frac{\lambda_i}{c_i}\int\mathrm{d}^3\br\;\mathrm{d}^3\bb\;e^{-w_i(\br,\bb)-\beta U_i(\bb)+\epsilon_i}\\
&&\hspace{2.1cm}\times e^{i\int\mathrm{d}^3\br'\hat n_i(\br';\br,\bb)\left[\phi(\br')+i f\ce(\br')\right]}.\nonumber
\eea
In Eq.~(\ref{e5}), the first term is the electrostatic free energy of the implicit solvent, the second term is the contribution from the fixed macromolecular charges, and the third term corresponds to the fluctuating density of the solute molecules. In the present work, we consider two different solute models introduced below.

\subsubsection{Linear solute molecules}

Fig.~\ref{fig1}(b) illustrates the linear solute model accounting for the charge spread and flexibility of extended polyamines, proteins, and biomolecules, such as putrescine~\cite{putr}, peptides~\cite{pep}, avidin~\cite{avid}, zwitterionic molecules~\cite{zwit1,zwit2}, and microswimmers~\cite{microsw1}.  The corresponding solute geometry is equally relevant to the understanding of the ionic polarizability effects commonly taken into account with a Drude oscillator potential~\cite{DrudeMD1,DrudeMD2,DrudeMD3,DrudeMD4}. The plot shows that each solute molecule carries the C.M. charge  $p_i$ and the second terminal charge $e_i$ of arbitrary sign. These charges are separated by the fluctuating distance $\bb$. Thus, the charge structure function of the molecule with valency $q_i=p_i+e_i$ reads
\be\label{cs}
\hat n_i(\br';\br,\bb)=p_i\delta^3(\br'-\br)+e_i\delta^3(\br'-\br-\bb).
\ee
Moreover, the flexibility of these molecules is incorporated via a Drude oscillator potential~\cite{drude} 
\be\label{dr}
\beta U_i(\bb)=\frac{(b-a_i)^2}{\alpha^2_i}.
\ee
The potential~(\ref{dr}) accounts for the fluctuations of the solute size $b$ around its average $a_i$ with the variance $\alpha_i$~\cite{rem0}. 

Together with Eqs.~(\ref{cs})-(\ref{dr}), the Hamiltonian~(\ref{e5}) takes for $f\ce(\br)=0$ the specific form
\bea
\label{e5II}
\hspace{0mm}\beta H[\phi]&=&\frac{k_{\rm B}T}{2e^2}\int\mathrm{d}^3\br\;\e(\br)\left[\nabla\phi(\br)\right]^2-i\int\mathrm{d}^3\br\sigma(\br)\phi(\br)\nonumber\\
&&-\sum_{i=1}^s\lambda_i\int\frac{\mathrm{d}^3\bb}{(\sqrt{\pi}\alpha_i)^3c_i}e^{-\frac{(b-a_i)^2}{\alpha^2_i}}\\
&&\hspace{8mm}\times\int\mathrm{d}^3\br\;e^{-w_i(\br,\bb)+\epsilon_i}e^{i\left[p_i\phi(\br)+e_i\phi(\br+\bb)\right]}.\nonumber
\eea
In Eq.~(\ref{e5II}), the normalization factor $c_i$ of the probability distribution characterizing the solute fluctuations is defined as $c_i=\int{\rm d}^3\bb\;e^{-\beta U_i(b)}/(\sqrt{\pi}\alpha_i)^3$, or
\be
\label{ci}
c_i=\frac{2}{\sqrt\pi}\frac{a_i}{\alpha_i}e^{-\frac{a_i^2}{\alpha_i^2}}+\left(1+\frac{2a_i^2}{\alpha_i^2}\right)\left[1+{\rm erf}\left(\frac{a_i}{\alpha_i}\right)\right],
\ee
where we used the error function ${\rm erf}(x)$~\cite{math}. In the limit of a vanishing average size ($a_i\to0$), the coefficient~(\ref{ci}) tends to unity ($c_i\to1$).

\subsubsection{Spherical solute molecules}

The second type of solute model consists of the spherical charge distribution depicted in Fig.~\ref{fig1}(c). This particular solute structure represents the geometry of  micelles, bilayer vesicles and liposomes, and colloids dressed by surfactants~\cite{Isr,sp0,sp1,sp2}. The charge confinement to the spherical solute volume of radius $b$ implies a charge structure function of the form 
\be\label{cs2}
\hat{n}_i(\br';\br,\bb)=\hat{n}_i\left(||\br'-\br||,b\right). 
\ee
The isotropic thermal fluctuations of these spherical molecules will be incorporated with a modified Drude potential defined in terms of the Boltzmann distribution
\be
\label{dr2}
e^{-\beta U_i(\bb)}=e^{-\frac{b^2}{\alpha^2_i}}\theta(b-b_0).
\ee
In Eq.~(\ref{dr2}), the Heaviside function $\theta(x)$~\cite{math} imposing the minimum radial size $b_0<d/2$ accounts for the rigidity of the molecule in its shrunken configuration. Incorporating these features into Eq.~(\ref{e5}), the Hamiltonian becomes
\bea
\label{hs}
\hspace{0mm}\beta H[\phi]&=&\frac{k_{\rm B}T}{2e^2}\int\mathrm{d}^3\br\;\e(\br)\left[\nabla\phi(\br)\right]^2-i\int\mathrm{d}^3\br\sigma(\br)\phi(\br)\nonumber\\
&&-\sum_{i=1}^s\lambda_i\int_{b_0}^\infty\frac{\mathrm{d}b}{\sqrt{\pi}\alpha_ic_i}e^{-\frac{b^2}{\alpha^2_i}}\int\mathrm{d}^3\br\;e^{-w_i(\br,\bb)+\epsilon_i}\nonumber\\
&&\hspace{2.5cm}\times e^{i\int\mathrm{d}^3\br'\hat{n}_i(||\br'-\br||,b)\phi(\br')},
\eea
where we introduced the normalization factor 
\be\label{f2}
c_i=\frac{1}{2}\left[1-{\rm erf}\left(\frac{b_0}{\alpha_i}\right)\right].
\ee

\subsection{Generalized contact-value theorem} 

The solvent-implicit contact value theorem relating the intermembrane pressure to the contact density of the salt ions has been previously derived for point-like charges~\cite{NetzCont,MoreiraCont,DavidCont}. In this part, we generalize this exact identity to the case of structured solute molecules in dielectrically homogeneous systems. 

The interplate pressure follows from the variation of the grand potential $\beta\Omega_{\rm G}=-\ln Z_{\rm G}$ with respect to the intermembrane distance at fixed solute fugacity,
\be
\label{pr1}
\beta P=-\frac{\delta(\beta\Omega_{\rm G})}{\delta(Sd)}=-\frac{1}{S}\lan\frac{\delta\left(\beta H\right)}{\delta d}\ran,
\ee
where the field average of the general functional $F[\phi(\br)]$ is defined as
\be
\label{fieldav}
\lan F[\phi(\br)]\ran=\frac{1}{Z_{\rm G}}\int\frac{\mathcal{D}\phi}{\sqrt{{\rm det}\left(v\ce\right)}}\;e^{-\beta H[\phi]}F[\phi(\br)].
\ee
For the explicit evaluation of Eq.~(\ref{pr1}), we will follow the approach of Refs.~\cite{NetzCont,MoreiraCont} that consists in recasting the Hamiltonian functional~(\ref{e5}) in a form that facilitates the treatment of the membrane charge term embodying the interfacial discontinuity of the electric field. To this aim, we set $f\ce(\br)=0$, and reexpress the partition function~(\ref{par}) and the Hamiltonian~(\ref{e5}) in terms of the fluctuating potential $\psi(\br)=\phi(\br)-\phi\ce(\br)$ shifted by the membrane charge-induced bare potential 
\be\label{ap}
\phi\ce(\br)=i\int\mathrm{d}^3\br'v\ce(\br,\br')\sigma(\br')
\ee
as $Z_{\rm G}=\int\mathcal{D}\psi\;e^{-\beta H[\psi]}/\sqrt{{\rm det}\left(v\ce\right)}$, and 
\bea
\label{e5III}
\beta H[\psi]&=&\int\frac{\mathrm{d}^3\br\mathrm{d}^3\br'}{2}\psi(\br)v\ce^{-1}(\br,\br')\psi(\br')\\
&&+\int\frac{\mathrm{d}^3\br\mathrm{d}^3\br'}{2}\sigma(\br)v\ce(\br,\br')\sigma(\br')\nonumber\\
&&-\sum_{i=1}^s\frac{\lambda_i}{c_i}\int\mathrm{d}^3\br\;\mathrm{d}^3\bb\;e^{-w_i(\br,\bb)-\beta U_i(\bb)+\epsilon_i}\nonumber\\
&&\hspace{2cm}\times e^{i\int\mathrm{d}^3\br'\hat n_i(\br';\br,\bb)\left[\psi(\br')+\phi\ce(\br')\right]}.\nonumber
\eea
Evaluating now the derivative of Eq.~(\ref{e5III}) with respect to the intermembrane distance $d$, and restoring the original potential $\phi(\br)$, the pressure~(\ref{pr1}) follows in the form
\begin{widetext}
\bea
\label{pr2}
\beta P&=&-\frac{1}{S} \frac{\delta}{\delta d}\left\{\int\frac{\mathrm{d}^3\br\mathrm{d}^3\br'}{2}\sigma(\br)v\ce(\br,\br')\sigma(\br')\right\}\\
&&+\frac{1}{S}\sum_{i=1}^s\frac{\lambda_i}{c_i}\int\mathrm{d}^3\br\;\mathrm{d}^3\bb\;e^{-\beta U_i(\bb)+\epsilon_i}\lan e^{i\int\mathrm{d}^3\br'\hat n_i(\br';\br,\bb)\phi(\br')}\ran
\left\{\frac{\delta}{\delta d}e^{-w_i(\br,\bb)}+ie^{-w_i(\br,\bb)}\int\mathrm{d}^3\br''\hat n_i(\br'';\br,\bb)\frac{\delta \phi\ce(\br'')}{\delta d}\right\}.\nonumber
\eea

Next, we insert into Eq.~(\ref{ap}) and the first term on the r.h.s. of Eq.~(\ref{pr2}) the Fourier transform of the Coulomb potential $v\ce(r)=2\pi\ell_{\rm B}\int\mathrm{d}\bk\;e^{i\bk\cdot\br_\pa-k|z|}/(4\pi^2k)$, where $\br_\pa$ stands for the component of the position vector $\br=\br_\pa+z\hat u_z$ along the membrane surfaces. After some simple algebra, Eq.~(\ref{pr2}) simplifies to
\bea
\label{pr3}
\beta P&=&2\pi\ell_{\rm B}\sigma_{\rm m}^2\\
&&+\frac{1}{S}\sum_{i=1}^s\frac{\lambda_i}{c_i}\int\mathrm{d}^3\br\;\mathrm{d}^3\bb\;e^{-\beta U_i(\bb)+\epsilon_i}\lan e^{i\int\mathrm{d}^3\br'\hat n_i(\br';\br,\bb)\phi(\br')}\ran
\left\{\frac{\delta}{\delta d}e^{-w_i(\br,\bb)}-2\pi\ell_{\rm B}\sigma_{\rm m} e^{-w_i(\br,\bb)}\int\mathrm{d}^3\br''\hat n_i(\br'';\br,\bb)\right\}.\nonumber
\eea
\end{widetext}
In Eq.~(\ref{pr3}), the first term on the r.h.s. corresponds to the repulsive electrostatic coupling between the fixed membrane charges at $z=0$ and $z=d$. Then, the first term in the curly bracket is the repulsive osmotic contribution from the solute molecules to the intermembrane interactions. The latter is induced by the steric solute-membrane coupling mediated by the potential $w_i(\br,\bb)$. Finally, the second term in the bracket accounts for the electrostatic solute-membrane charge attraction. This term can be expressed in terms of the average charge density of the confined solute molecules given by 
\bea
\label{pr4}
\rho\ce(\br)&=&\lan\hat\rho\ce(r)\ran=-\frac{1}{Z_{\rm G}}\left.\frac{\delta Z_{\rm G}}{\delta f\ce(\br)}\right|_{f\ce(r)=0}\nonumber\\
&=&\sum_{i=1}^s\frac{\lambda_i}{c_i}\int\mathrm{d}^3\br''\mathrm{d}^3\bb\;e^{-w_i(\br'',\bb)-\beta U_i(\bb)+\epsilon_i}\\
&&\hspace{1cm}\times\lan e^{i\int\mathrm{d}^3\br'\hat n_i(\br';\br'',\bb)\phi(\br')}\ran\hat n_i(\br;\br'',\bb).\nonumber
\eea
Using Eq.~(\ref{pr4}), Eq.~(\ref{pr3}) reduces to
\bea
\label{pr5}
\beta P&=&2\pi\ell_{\rm B}\sigma_{\rm m}^2-\frac{2\pi\ell_{\rm B}\sigma_{\rm m}}{S}\int\mathrm{d}^3\br\rho\ce(\br)\\
&&+\frac{1}{S}\sum_{i=1}^s\frac{\lambda_i}{c_i}\int\mathrm{d}^3\br\;\mathrm{d}^3\bb\;e^{-\beta U_i(\bb)+\epsilon_i}\frac{\delta}{\delta d}e^{-w_i(\br,\bb)}\nonumber\\
&&\hspace{3cm}\times\lan e^{i\int\mathrm{d}^3\br'\hat n_i(\br';\br,\bb)\phi(\br')}\ran.\nonumber
\eea
Taking now into account the electroneutrality condition $\int\mathrm{d}^3\br\rho\ce(\br)=2\sigma_{\rm m} S$, one finds that the second term on the r.h.s. of Eq.~(\ref{pr5}) corresponds to the standard attractive force $-4\pi\ell_{\rm B}\sigma_{\rm m}^2$ between the interfacial counterion layers and the membrane charges~\cite{NetzCont,MoreiraCont}. Consequently, the interaction pressure~(\ref{pr5}) becomes
\bea
\label{pr6}
\beta P&=&\frac{1}{S}\sum_{i=1}^s\frac{\lambda_i}{c_i}\int\mathrm{d}^3\br\;\mathrm{d}^3\bb\;e^{-\beta U_i(\bb)+\epsilon_i}\frac{\delta}{\delta d}e^{-w_i(\br,\bb)}\nonumber\\
&&\hspace{2.8cm}\times\lan e^{i\int\mathrm{d}^3\br'\hat n_i(\br';\br,\bb)\phi(\br')}\ran\nonumber\\
&&-2\pi\ell_{\rm B}\sigma_{\rm m}^2.
\eea

The generalized contact theorem~(\ref{pr6}) valid for any solute structure and electrostatic coupling strength indicates that the intramolecular solute composition alters exclusively the repulsive osmotic pressure component without affecting the electrostatic attraction force. Below, we evaluate this identity for linear and spherical solute molecules confined to adjacent membrane walls.

\subsubsection{Linear solute molecules}

In order to evaluate the contact theorem~(\ref{pr6}) in the specific case of linear solute molecules, we derive first the number densities of the terminal charges. To this aim, we specify the steric solute potential as 
\be\label{s1}
w_i(\br,\bb)=w_{i,p}(\br)+w_{i,e}(\br+\bb), 
\ee
where the first and second terms on the r.h.s. are the potentials acting on the C.M. charge and the neighbouring terminal charge on each solute molecule, respectively.  The average number densities of these charges follow from the identities $\rho_{i,\alpha}(\br)=\delta(\beta\Omega_{\rm G})/\delta w_{i,\alpha}(\br)=\lan\delta(\beta H)/\delta w_{i,\alpha}(\br)\ran$ for $\alpha=\{p,e\}$, with the dimensionless grand potential defined as $\beta\Omega_{\rm G}=-\ln Z_{\rm G}$, and the Hamiltonian given by Eq.~(\ref{e5II}). One obtains 
\bea
\label{e7}
\rho_{i,p}(\br)&=&\frac{\lambda_i}{c_i}\int\frac{\mathrm{d}^3\bb}{(\sqrt{\pi}\alpha_i)^3}e^{-\frac{(b-a_i)^2}{\alpha^2_i}-w_i(\br,\bb)+\epsilon_i}\nonumber\\
&&\hspace{2.2cm}\times\lan e^{i\left[p_i\phi(\br)+e_i\phi(\br+\bb)\right]}\ran;\\
\label{e8}
\rho_{i,e}(\br)&=&\frac{\lambda_i}{c_i}\int\frac{\mathrm{d}^3\bb}{(\sqrt{\pi}\alpha_i)^3}e^{-\frac{(b-a_i)^2}{\alpha^2_i}-w_i(\br-\bb,\bb)+\epsilon_i}\nonumber\\
&&\hspace{2.2cm}\times\lan e^{i\left[p_i\phi(\br-\bb)+e_i\phi(\br)\right]}\ran.
\eea

In order to account for the confinement of the solute charges to the intermembrane region, we choose the steric potentials in Eq.~(\ref{s1}) as 
\be
\label{st1}
e^{-w_{i,p}(z)}=e^{-w_{i,e}(z)}=\theta(z)\theta(d-z).
\ee
Plugging Eqs.~(\ref{cs}), (\ref{s1}), and~(\ref{st1}) into Eqs.~(\ref{e7})-(\ref{e8}), and accounting for the translational symmetry of the system along the membrane surfaces, which implies $\rho_{i,\alpha}(\br)=\rho_{i,\alpha}(z)$ for $\alpha=\{p,e\}$ , the terminal charge densities follow for $0<z<d$ in the form
\bea
\label{e7II}
\rho_{i,p}(z)&=&\frac{\lambda_i}{c_i}\int\frac{\mathrm{d}^2\bb_\pa}{(\sqrt{\pi}\alpha_i)^3}\int_{-z}^{d-z}\mathrm{d}b_ze^{-\frac{(b-a_i)^2)}{\alpha^2_i}+\epsilon_i}\\
&&\hspace{2.4cm}\times\lan e^{i\left[p_i\phi(\br)+e_i\phi(\br+\bb)\right]}\ran;\nonumber\\
\label{e8II}
\rho_{i,e}(z)&=&\frac{\lambda_i}{c_i}\int\frac{\mathrm{d}^2\bb_\pa}{(\sqrt{\pi}\alpha_i)^3}\int_{-(d-z)}^{z}\mathrm{d}b_ze^{-\frac{(b-a_i)^2}{\alpha^2_i}+\epsilon_i}\\
&&\hspace{2.4cm}\times\lan e^{i\left[p_i\phi(\br-\bb)+e_i\phi(\br)\right]}\ran.\nonumber
\eea
In Eqs.~(\ref{e7II})-(\ref{e8II}), $\bb_\pa$ and $b_z$ stand for the components of the molecular length vector $\bb$ parallel and normal to the membrane surfaces, respectively.

We combine now Eqs.~(\ref{s1}) and~(\ref{st1}) to obtain
\bea
\label{pr7}
\hspace{-1mm}\frac{\delta}{\delta d}e^{-w_i(\br,\bb)}&=&\theta(z)\theta(z+b_z)\left\{\theta(d-z-b_z)\delta(d-z)\right.\\
&&\hspace{2.1cm}\left.+\theta(d-z)\delta(d-z-b_z)\right\}.\nonumber
\eea
Inserting the structure factor~(\ref{cs}) together with Eq.~(\ref{pr7}) into the contact value identity~(\ref{pr6}), and identifying in the resulting expression the density~(\ref{e7II}) of the C.M. charge, the intermembrane pressure reduces to
\bea
\label{pr8}
\beta P&=&\sum_{i=1}^s\rho_{i,p}(d)-2\pi\ell_{\rm B}\sigma_{\rm m}^2\\
&&+\frac{1}{S}\sum_{i=1}^s\frac{\lambda_i}{c_i}\int\frac{\mathrm{d}^2\bb_\pa}{(\sqrt{\pi}\alpha_i)^3}\int\mathrm{d}^2\br_\pa\nonumber\\
&&\hspace{1cm}\times\int_0^d\mathrm{d}b_ze^{-\frac{(b-a_i)^2}{\alpha^2_i}+\epsilon_i}\nonumber\\
&&\hspace{2cm}\times\lan e^{i\left[p_i\phi(\br_\pa,d-b_z)+e_i\phi(\br_\pa+\bb_\pa,d)\right]}\ran.\nonumber
\eea
Finally, introducing in the integral term of Eq.~(\ref{pr8}) the change of variable $\br_\pa\to\br'_\pa=\br_\pa+\bb_\pa$, one finds that this term corresponds to the density ~(\ref{e8II}) of the terminal charge $e_i$. Thus, the contact value identity follows as
\be
\label{pr9}
\beta P=\sum_{i=1}^s\left[\rho_{i,p}(d)+\rho_{i,e}(d)\right]-2\pi\ell_{\rm B}\sigma_{\rm m}^2.
\ee

\subsubsection{Spherical solute molecules}

The C.M. density of the spherical solute molecules can be obtained by setting the steric potential to $w_i(\br,\bb)=w_i(\br)$, and evaluating the functional derivative $\rho_i(\br)=\delta(\beta\Omega_{\rm G})/\delta w_i(\br)=\lan \delta(\beta H)/\delta w_i(\br)\ran$ with the Hamiltonian~(\ref{hs}). This yields
\bea
\label{s0}
\rho_i(\br)&=&\frac{\lambda_i}{c_i}\int_{b_0}^{\infty}\frac{\mathrm{d}b}{\sqrt{\pi}\alpha_i}e^{-\frac{b^2}{\alpha^2_i}-w_i(\br,\bb)+\epsilon_i}\\
&&\hspace{2.cm}\times\lan e^{i\int\mathrm{d}^3\br'\hat{n}_i(||\br'-\br||,b)\phi(\br')}\ran.\nonumber
\eea
In order to determine the functional form of the steric potential $w_i(\br,\bb)$, one has to specify the steric constraints on the solute molecules. First, we note that the rigidity of the spherical molecule at its maximally compressed size $b=b_0$ means that $b_0$ is the closest approach distance of its C.M. to the membrane surfaces, i.e. $b_0\leq z\leq d-b_0$. Moreover, the impenetrability of the membrane implies that the fluctuations of the spherical boundary of the solute molecules are limited to the intermembrane region, i.e. $0\leq z-b$ and $z+b\leq d$. Combining these conditions, the steric constraint on the C.M. position and size of each solute molecule can be expressed as
\be
\label{s2}
e^{-w_i(\br,\bb)}=\theta(z-b_0)\theta(d-b_0-z)\theta(b-b_0)\theta\left[b_+(z)-b\right],
\ee
where we introduced the auxiliary function
\be
\label{au1}
b_+(z)={\rm min}(z,d-z).
\ee
Inserting now the steric potential~(\ref{s2}) into Eq.~(\ref{s0}), the C.M. density of the spherical molecules becomes
\bea
\label{s3}
\rho_i(\br)&=&\lambda_i\theta(z-b_0)\theta(d-b_0-z)\\
&&\times\int_{b_0}^{b_+(z)}\frac{\mathrm{d}b}{\sqrt{\pi}\alpha_ic_i}e^{-\frac{(b-a_i)^2}{\alpha^2_i}+\epsilon_i}\nonumber\\
&&\hspace{2.2cm}\times\lan e^{i\int\mathrm{d}^3\br'\hat{n}_i(||\br'-\br||,b)\phi(\br')}\ran.\nonumber
\eea

In order to calculate the pressure~(\ref{pr6}), one needs to evaluate the derivative of Eq.~(\ref{s2}) with respect to the intermembrane distance. This yields
\be
\label{s4}
\frac{\delta}{\delta d}e^{-w_i(\br,\bb)}=\theta(z-d/2)\theta(d-b_0-z)\delta(b-d+z).
\ee
Injecting Eqs.~(\ref{cs2})-(\ref{dr2}) and~(\ref{s4}) into the contact theorem~(\ref{pr6}), the intermembrane force follows in the form
\bea
\label{s5}
\beta P&=&\frac{1}{S}\sum_{i=1}^s\lambda_i\int\mathrm{d}^2\br_\pa\int_{b_0}^{d/2}\frac{\mathrm{d}b}{\sqrt{\pi}\alpha_ic_i}e^{-\frac{b^2}{\alpha^2_i}+\epsilon_i}\\
&&\hspace{1.8cm}\times\left.\lan e^{i\int\mathrm{d}^3\br'\hat{n}_i(||\br'-\br||,b)\phi(\br')}\ran\right|_{z=d-b}\nonumber\\
&&-2\pi\ell_{\rm B}\sigma_{\rm m}^2\nonumber
\eea

Due to the diffuse inner structure of the spherical solute particles, the pressure~(\ref{s5})  cannot be expressed in terms of the solute density~(\ref{s3}). However, the intramolecular charge configurations embodied by the repulsive component of the pressure~(\ref{s5}) can be identified via the constraint~(\ref{s4}). Indeed, one notes that this identity imposes on the C.M. coordinate $z$ the constraints $z+b=d$ and $d/2<z<d-b_0$.  Hence, the integral term of Eq.~(\ref{s5}) incorporating the effect of the solute-wall collisions corresponds to the trace over the spherical charge configurations characterized by the solute boundary in contact with the membrane wall, and the C.M. of the molecule oscillating between the mid-pore and the closest approach distance $d-b_0$. 

\section{MF regime of macromolecular interactions in structured liquids}
\label{res}

Owing to the breaking of the planar membrane symmetry by the spherical charge geometry, the MF-level electrostatic equation of state for spherical ions corresponds to a three dimensional integro-differential equation. As the numerical solution of this equation is a formidable computational task beyond the scope of the present work, we investigate here MF-level intermembrane interactions exclusively in the case of a 1:1 electrolyte composed of the linear solute molecules depicted in Fig.~\ref{fig1}(b). In Section~\ref{spm}, the case of spherical solute charges will be considered in the electrostatic SC regime where the vanishing intermembrane field enables the fully analytical characterization of the membrane interactions.

The MF approximation ignoring many-body interactions also neglects the intramolecular charge interactions. Thus, we set $\epsilon_i=0$. Subtracting from Eq.~(\ref{pr9}) the bulk osmotic pressure $\beta P_{\rm b}=\sum_{i=1}^s\rho_{i{\rm b}}$ acting on the outer membrane walls, the net pressure follows as 
\be\label{net}
\beta\pn=\sum_{i=1}^s\left[\rho_{i,p}(d)+\rho_{i,e}(d)-\rho_{i{\rm b}}\right]-2\pi\ell_{\rm B}\sigma_{\rm m}^2.
\ee
 In Appendix~\ref{cont}, the MF limit of Eq.~(\ref{net}) is derived directly from the MF grand potential of the system.
 
Within the framework of the PB formalism for point charges, the contact theorem relating the MF contact density of the ions and the pressure $\pn^{\rm (PB)}$ is given by
\be
\label{cpb}
\beta\pn^{\rm (PB)}=\sum_{i=1}^s\left[\rho_i(d)-\rho_{i{\rm b}}\right]-2\pi\ell_{\rm B}\sigma_{\rm m}^2,
\ee
where $\rho_i(z)=\rho_{i{\rm b}}e^{-q_i\phi^{\rm(PB)}(z)}$ is the MF ion density, and $\phi^{\rm(PB)}(z)$ is the potential solving the PB Eq.~\cite{Isr}. The comparison of the contact theorem~(\ref{net}) with its PB counterpart~(\ref{cpb}) indicates that the main peculiarity of the former is the separate pressure contribution from each terminal solute charge. In Secs.~\ref{ch} and~\ref{lsc}, this feature will be shown to have a major impact on the short-range branch of the intermembrane repulsion.

\subsection{MF-level electrostatic equation of state}
\label{MFpr}

In this section, solute structure effects will be characterized via the comparison of the PB pressure~(\ref{cpb}) and the generalized contact theorem~(\ref{net}) computed with the MF limit of the terminal charge densities~(\ref{e7II})-(\ref{e8II}). In the MF approximation, the electrostatic potential required for the evaluation of these charge densities satisfies the saddle-point condition $\delta(\beta H)/\delta\phi(\br)=0$ of the Hamiltonian~(\ref{e5II}). Passing from the complex to the real potential via the transformation $\phi(\br)\to i\phi(\br)$, this condition follows as an extended PB equation, 
\bea\label{e6}
&&\frac{k_{\rm B}T}{e^2}\nabla\cdot\e(\br)\nabla\phi(\br)+\sigma(\br)\\
&&+\sum_{i=1}^s\frac{\lambda_i}{c_i}\int\frac{\mathrm{d}^3\bb}{(\sqrt{\pi}\alpha_i)^3}
\left\{p_ie^{-w_i(\br,\bb)-\Phi_{i,p}(\br,\bb)}\right.\nonumber\\
&&\hspace{3.5cm}\left.+e_ie^{-w_i(\br-\bb,\bb)-\Phi_{i,e}(\br,\bb)}\right\}=0,\nonumber
\eea
with the potential energies experienced by the terminal charges $p$ and $e$ given by
\bea
\label{p1}
\hspace{-5mm}\Phi_{i,p}(\br,\bb)&=&\frac{(b-a_i)^2}{\alpha^2_i}+p_i\phi(\br)+e_i\phi(\br+\bb);\\
\label{p2}
\hspace{-5mm}\Phi_{i,e}(\br,\bb)&=&\frac{(b-a_i)^2}{\alpha^2_i}+p_i\phi(\br-\bb)+e_i\phi(\br).
\eea

At the electrostatic MF-level, the statistical average in Eq.~(\ref{fieldav}) simplifies as $\lan F[\phi]\ran=F[\phi]$, where $\phi(\br)$ is the solution of the saddle-point Eq.~(\ref{e6}). By expressing the charge densities~(\ref{e7})-(\ref{e8}) in this MF approximation, taking the bulk reservoir limit where the electrostatic potential $\phi(\br)$ and the steric potentials $w_{i,\alpha}(\br)$ vanish, and evaluating the integrals over the intramolecular fluctuations, one finds that the fugacity of each molecular species equals its bulk concentration, i.e. $\lambda_i=\rho_{i{\rm b}}$. Moreover, the translational invariance of the charge distribution along the membrane walls implies the presence of an exclusively perpendicular average field, i.e. $\phi(\br)=\phi(z)$. Consequently, Eq.~(\ref{e6}) takes the form of a one-dimensional integro-differential Poisson Eq.,
\bea
\label{e11}
&&\partial_z^2\phi(z)+4\pi\ell_{\rm B}\sum_{i=1}^s\left[p_i\rho_{i,p}(z)+e_i\rho_{i,e}(z)\right]\\
&&=4\pi\ell_{\rm B}\sigma_{\rm m}\left[\delta(z)+\delta(d-z)\right].\nonumber
\eea
In Eq.~(\ref{e11}), the MF-level number densities of the terminal charges follow from Eqs.~(\ref{e7II})-(\ref{e8II}) as
\bea
\label{e12}
\rho_{i,p}(z)&=&\rho_{i{\rm b}}\int_{-z}^{d-z}\mathrm{d}b_z\;g_i(b_z)e^{-p_i\phi(z)-e_i\phi(z+b_z)};\\
\label{e12II}
\rho_{i,e}(z)&=&\rho_{i{\rm b}}\int_{-z}^{d-z}\mathrm{d}b_z\;g_i(b_z)e^{-p_i\phi(z+b_z)-e_i\phi(z)},
\eea
where the conditional probability for the intramolecular solute fluctuations normal to the membranes reads
\be
\label{gi}
g_i(b_z)=\frac{1}{\sqrt{\pi}\alpha_i}\frac{e^{-\frac{\left(|b_z|-a_i\right)^2}{\alpha_i^2}}+\sqrt{\pi}\frac{a_i}{\alpha_i}\left[1+{\rm erf}\left(\frac{a_i-|b_z|}{\alpha_i}\right)\right]}{\frac{2}{\sqrt\pi}\frac{a_i}{\alpha_i}e^{-\frac{a_i^2}{\alpha_i^2}}+\left(1+\frac{2a_i^2}{\alpha_i^2}\right)\left[1+{\rm erf}\left(\frac{a_i}{\alpha_i}\right)\right]}.
\ee
In the remainder, the MF-level pressure will be calculated by inserting the surface value of the charge densities~(\ref{e12})-(\ref{e12II}) into the contact value identity~(\ref{net}).

\begin{figure*}
\includegraphics[width=1.\linewidth]{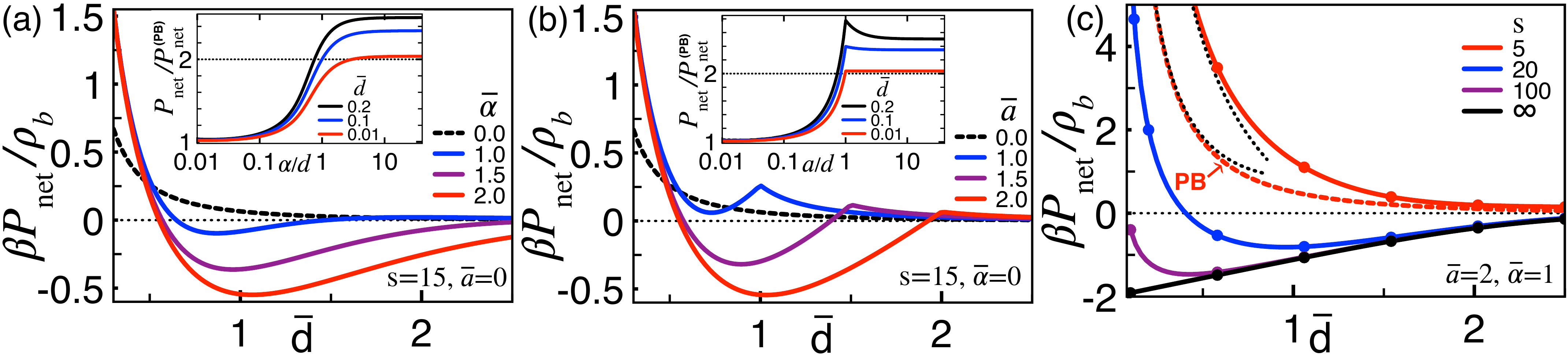}
\caption{(Color online) Main plots: interaction pressure~(\ref{net}) versus the intermembrane distance for (a) different solute flexibilities and vanishing average size, (b) various rigid solute sizes ($\alpha=0$), and (c) different values of the parameter $s=\kappa\mu\propto\sqrt{\rho_{\rm b}}/\sigma_{\rm m}$ (solid curves). In (c), the circles mark the Donnan formula~(\ref{e25}),  and the dotted black lines display for $s=5$ the short-distance asymptotic laws~(\ref{e32}) and~(\ref{e33}). The dashed curves in (a)-(c) correspond to the PB pressure~(\ref{cpb}) ($\bar{a}=\bar{\alpha}=0$). The insets in (a)-(b) illustrate the ratio of the net pressure and its PB limit against the  solute polarizability and size at short distances.} 
\label{fig2}
\end{figure*}

From now on, we will refer to Eq.~(\ref{e11}) as the structured PB (SPB) equation. The rigid rod case studied in Refs.~\cite{pincus,bohinc2} follows in the limit $\alpha_i\to0$ of Eq.~(\ref{gi}) as  
\be
\label{gi2}
\lim_{\alpha_i\to0}g_i(b_z)=\frac{1}{2a_i}\theta\left(|b_z|-a_i\right).
\ee
Then, if the fixed average size vanishes, the probability function~(\ref{gi}) reduces to a gaussian distribution, i.e.
\be
\label{gi3}
\lim_{a_i\to0}g_i(b_z)=\frac{1}{\sqrt{\pi}\alpha_i}e^{-\frac{b_z^2}{\alpha_i^2}}.
\ee

In this work, the exact numerical solution of the SPB Eq.~(\ref{e11}) was carried out via a recursive algorithm. The details of this numerical scheme are reported in Appendix~\ref{nums}. The boundary conditions required for this solution follows from the integration of Eq.~(\ref{e11}) in the vicinity of the membrane surfaces at $z=0$ and $z=d$, 
\be
\label{gauss}
\phi'(0^+)=2/\mu\;;\hspace{5mm}\phi'(d^-)=-2/\mu,
\ee
where we used the Gouy-Chapman length 
\be\label{gc}
\mu=\left(2\pi |q_i|\ell_{\rm B}\sigma_{\rm m}\right)^{-1}. 
\ee
We also note that the bulk electroneutrality condition in the reservoir follows from the bulk limit of Eq.~(\ref{e11}) as
\be
\label{e12III}
\sum_{i=1}^s\rho_{i{\rm b}}q_i=0.
\ee

\subsection{Effect of the extended charge structure on the intermembrane interactions}
\label{ch}

The symmetric electrolyte is composed of two solute species ($s=2$), each carrying opposite terminal charges. The latter are set to $p_-=1$, $e_-=-2$, $p_+=2$, and $e_+=-1$, where  the indices $i=+$ and $i=-$ refer  to the cationic and anionic solute species, respectively. Together with the electroneutrality condition~(\ref{e12III}), this yields the net solute valencies $q_\pm=\pm1$ and salt concentrations $\rho_{\pm{\rm b}}=\rho_{\rm b}$. Moreover, with the aim to simplify the analysis, both solute species are assumed to have the same average size ($a_i=a$) and flexibility ($\alpha_\pm=\alpha$).  

In the main plots of Figs.~\ref{fig2}(a)-(b), we plotted the pressure~(\ref{net}) versus the distance $\bar{d}=\kappa d$ rescaled by the inverse Debye-H\"{u}ckle (DH) length $\kappa=\sqrt{8\pi\ell_{\rm B}\rho_{i{\rm b}}}$ at the dimensionless parameter $s=\kappa\mu=15$.  These plots show that the extended inner structure of the confined solute molecules gives rise to the coexistence of a large-distance attractive regime and a short-distance repulsive force of enhanced magnitude. Namely, Fig.~\ref{fig2}(a) indicates that upon the rise of the flexibility $\bar{\alpha}=\kappa\alpha$ at vanishing average size ($\bar{a}=\kappa a=0$), the large distance branch of the PB pressure (dashed curve) turns from weakly repulsive to substantially attractive. However, at short distances, the repulsive PB pressure rises with the molecular flexibility and saturates at $\alpha\sim d$. Finally, Fig.~\ref{fig2}(b) shows that in the case of purely rigid molecules ($\alpha=0$), the increase of the fixed solute size $\bar{a}$ causes a similar alteration of the PB pressure, with the additional emergence of a cusp at the separation distance $d=a$ where the rotational entropy of the confined molecules rises significantly.

With the aim to identify the effect of the salt concentration $\rho_{\rm b}$ and the membrane charge density $\sigma_{\rm m}$ on these features, in Fig.~\ref{fig2}(c), we reported the pressure at various values of the dimensionless parameter $s\propto\sqrt{\rho_{\rm b}}/\sigma_{\rm m}$. One sees that upon the increase of this parameter, the attractive pressure regime overrides gradually the shorter-range repulsive branch. Above the characteristic value $s\sim100$, the interaction force becomes purely attractive. Hence, salt increment or the reduction of the membrane charge density drives the system from the solute size-enhanced pure repulsion to the intermembrane attraction regime. 

\subsubsection{Long-range depletion attraction}

In order to identify the origin of the membrane attraction, we consider the neutral membrane limit $\sigma_{\rm m}=0$ (or $s\to\infty$) where the potential solving Eq.~(\ref{e11}) vanishes ($\phi(z)=0$). As a result, the densities~(\ref{e12})-(\ref{e12II}) reduce to 
\be\label{e27}
\rho_{i,p}(z)=\rho_{i,e}(z)=\rho_{\rm b}\gamma_i(z),
\ee
where we introduced the steric exclusion function $\gamma_i(z)=\int_{-z}^{d-z}\mathrm{d}b_zg_i(b_z)$. Evaluating the integral, one obtains
\bea\label{e17}
&&\hspace{-3mm}\gamma_i(z)=\frac{1}{c_i}\left\{\left(1+\frac{2a^2}{\alpha^2}\right){\rm erf}\left(\frac{a}{\alpha}\right)+\frac{ad}{\alpha^2}\right.\\
&&\hspace{13mm}+\left[\frac{1}{2}-\frac{a^2}{\alpha^2}\left(\frac{z}{a}-1\right)\right]{\rm erf}\left(\frac{z-a}{\alpha}\right)\nonumber\\
&&\hspace{13mm}+\left[\frac{1}{2}-\frac{a^2}{\alpha^2}\left(\frac{d-z}{a}-1\right)\right]{\rm erf}\left(\frac{d-z-a}{\alpha}\right)\nonumber\\
&&\hspace{13mm}\left.-\frac{a}{\sqrt{\pi}\alpha}\left[e^{-\frac{(d-z-a)^2}{\alpha^2}}+e^{-\frac{(z-a)^2}{\alpha^2}}-2e^{-\frac{a^2}{\alpha^2}}\right]\right\}\nonumber.
\eea
Consequently, the pressure~(\ref{net}) becomes
\be
\label{e28}
\beta\pn=-2\rho_{\rm b}\frac{\left[1-\frac{2a(d-a)}{\alpha^2}\right]{\rm erfc}\left(\frac{d-a}{\alpha}\right)+\frac{2a}{\sqrt{\pi}\alpha}e^{-\frac{(d-a)^2}{\alpha^2}}}{\frac{2}{\sqrt\pi}\frac{a}{\alpha}e^{-\frac{a^2}{\alpha^2}}+\left(1+\frac{2a^2}{\alpha^2}\right)\left[1+{\rm erf}\left(\frac{a}{\alpha}\right)\right]},
\ee
where ${\rm erfc}(x)$ is the complementary error function\cite{math}. 

\begin{figure}
\includegraphics[width=1\linewidth]{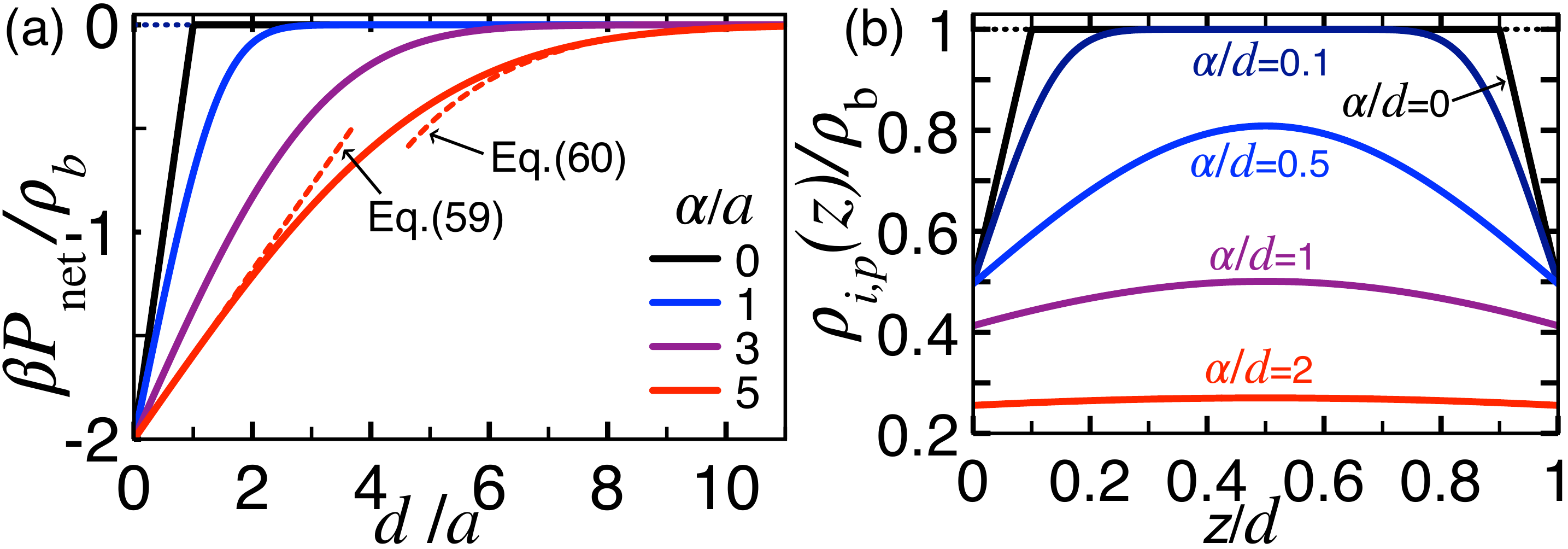}
\caption{(Color online) Interaction force and solute configuration between neutral membranes ($\sigma_{\rm m}=0$). (a) Pressure~(\ref{e28}) versus the intermembrane distance, and (b) ion density profile~(\ref{e27}) at the ion size $a=0.1\;d$ and various polarizabilities.}
\label{fig3}
\end{figure}

Fig.~\ref{fig3}(a) illustrates the purely attractive pressure~(\ref{e28}). In the case of rigid molecules (black curve for $\alpha=0$), the range of the linearly decaying pressure
\be\label{g3}
\lim_{\alpha\to0} \beta\pn=-2\rho_{\rm b}\left(1-\frac{d}{a}\right)\theta(a-d)
\ee
is set solely by the average solute size $a$. Then, in the presence of finite flexibility ($\alpha>0$), the intermembrane pressure acquiring an enhanced magnitude and range decays over the characteristic separation distance $d\sim\alpha$.  More precisely, at short separation distances, the interaction force~(\ref{e28}) decays linearly for $a<d\ll\alpha$ as
\be
\label{g7I}
\beta\pn=-2\rho_{\rm b}\left\{1-\frac{2}{\sqrt\pi}\frac{d}{\alpha}\left[1-\frac{4-\pi}{\sqrt\pi}\frac{a}{\alpha}\right]\right\}+O\left(\frac{d^2}{\alpha^2},\frac{a^2}{\alpha^2}\right).
\ee
Then, at large distances $a<\alpha\ll d$, the net pressure decreases according to a gaussian law,
\be
\label{g8I}
\beta\pn\approx-\frac{2\rho_{\rm b}}{\sqrt\pi}\frac{\alpha}{d}\left\{1+\frac{a}{d}\left[1-\frac{4}{\sqrt\pi}\frac{d}{\alpha}+\frac{2d^2}{\alpha^2}\right]\right\}e^{-\frac{d^2}{\alpha^2}}.
\ee
The asymptotic laws~(\ref{g7I})-(\ref{g8I}) are reported in Fig.~\ref{fig3}(a).

In Fig.~\ref{fig3}(b), the depletion mechanism driving the membrane attraction is illustrated in terms of the solute density~(\ref{e27}). In the case of purely rigid molecules (black curve for $\alpha=0$), the solute density reducing to
\be\label{g2}
\lim_{\alpha\to0}\gamma_i(z)=\frac{1}{2a}\left[{\rm min}(a,d-z)+{\rm min}(a,z)\right]
\ee
exhibits interfacial solute exclusion solely over the distance $a$. However, for flexible molecules, the entropic constraint limiting the solute fluctuations to the membrane separation distance causes partial solute exclusion from the entire intermembrane region. For $\alpha\gtrsim d$, this reduces the inner contact density of the terminal solute charges significantly below their outer contact density ($\gamma_i(d)<1/2$). As a result, the outer osmotic pressure on the membranes exceeds the inner pressure, driving the net force~(\ref{net}) into the attraction regime of Eq.~(\ref{e28}). 

\subsubsection{Short-range enhanced repulsion}
\label{don}

Within a uniform Donnan potential approximation, we characterize now the enhancement of the short-range repulsion in Fig.~\ref{fig2}. First, accounting for the plane symmetry of the system, the MF grand potential density obtained from the Hamiltonian~(\ref{e5}) as  $f=\beta H/S$ reads
\bea
\label{e13}
f&=&-\int_{-\infty}^\infty\frac{\mathrm{d}z}{8\pi\ell_{\rm B}}\left[\phi'(z)\right]^2-\sigma_{\rm m}\left[\phi(0)+\phi(d)\right]\\
&&-\sum_{i=1}^s\int_0^d\mathrm{d}z\rho_{i,p}(z).\nonumber
\eea
Substituting now the Donnan potential approximation $\phi(z)\approx\pd$ into the C.M. density~(\ref{e12}), the MF grand potential~(\ref{e13}) reduces to
\be
\label{e20}
f\approx-2\sigma_{\rm m}\pd-2d\rho_{\rm b}t_i\cosh(\pd),
\ee
with the partition coefficient $t_i=\int_0^d\mathrm{d}z\gamma_i(z)/d$, or
\bea
\label{e21}
t_i&=&\frac{1}{c_i}\left\{\frac{ad}{\alpha^2}+\left[1+\frac{a}{2\alpha^2d}(4da-2a^2-3\alpha^2)\right]{\rm erf}\left(\frac{a}{\alpha}\right)\right.\nonumber\\
&&\hspace{8mm}-\left[\frac{a}{\alpha^2d}(d-a)^2+\frac{3a}{2d}-1\right]{\rm erf}\left(\frac{d-a}{\alpha}\right)\nonumber\\
&&\hspace{8mm}+\frac{1}{\sqrt{\pi}\alpha d}\left[(a^2+\alpha^2-ad)e^{-\frac{(d-a)^2}{\alpha^2}}\right.\nonumber\\
&&\hspace{24mm}\left.-\left.(a^2+\alpha^2-2ad)e^{-\frac{a^2}{\alpha^2}}\right]\right\}.
\eea
Evaluating the saddle-point condition $\partial f/\partial\pd=0$ for Eq.~(\ref{e20}), one obtains the second order algebraic equation $e^{-\pd}-e^{\pd}-2r=0$. The solution of this equation yields 
\be
\label{e23}
\pd=-\ln\left(r+\sqrt{r^2+1}\right),
\ee
where the ratio of the membrane charge and solute charge densities has been defined as $r=\sigma_{\rm m}/(d\rho_{\rm b}t_i)$. 

\begin{figure}
\includegraphics[width=1.\linewidth]{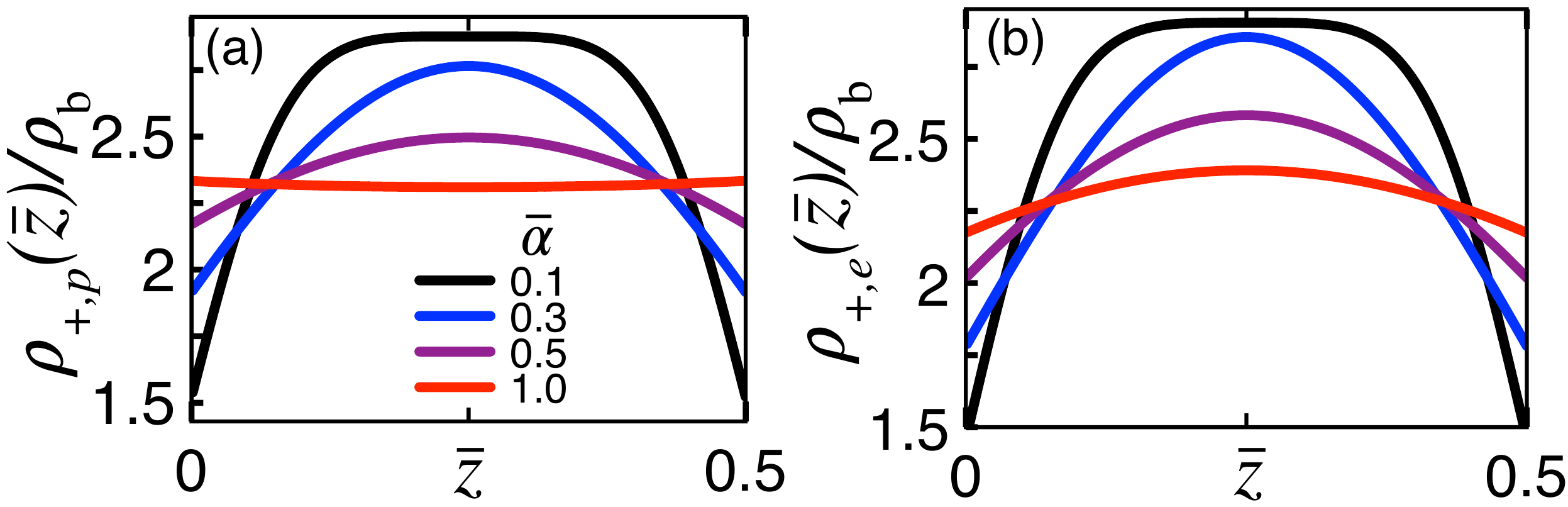}
\caption{(Color online) (a) Positive and (b) negative terminal charge densities~(\ref{e12})-(\ref{e12II}) of the cationic solute species. The intermembrane distance is $\bar{d}=0.5$, the dimensionless charge strength is $s=7$, and the average solute size is $a=0$. The values of the variance $\alpha$ are provided in the legend of (a).}
\label{fig4}
\end{figure}

Evaluating now the inner pressure $\beta P=-\delta f/\delta d$ with Eqs.~(\ref{e20})-(\ref{e23}), and subtracting the bulk pressure $\beta P_{\rm b}=-2\rho_{\rm b}$, the net pressure follows in the analytical form
\bea
\label{e25}
\beta\pn&=&2\rho_{\rm b}\left(\eta\sqrt{1+r^2}-1\right),
\eea
with the density renormalization factor of entropic origin,
\bea
\label{e25II}
\eta&=&\frac{1}{c_i}\left\{\frac{2ad}{\alpha^2}+\frac{2a}{\sqrt{\pi}\alpha}\left[e^{-\frac{a^2}{\alpha^2}}-e^{-\frac{(d-a)^2}{\alpha^2}}\right]\right.\\
&&\hspace{6mm}+\left(1+\frac{2a^2}{\alpha^2}\right){\rm erf}\left(\frac{a}{\alpha}\right)\nonumber\\
&&\hspace{6mm}\left.+\left[1-\frac{2a(d-a)}{\alpha^2}\right]{\rm erf}\left(\frac{d-a}{\alpha}\right)\nonumber\right\}.
\eea
Finally, taking the point charge limit $a\to0$ and $\alpha\to0$ of Eq.~(\ref{e25}), the Donnan approximation for the PB pressure follows as
\be\label{dpb}
\beta\pn^{({\rm PB})}=2\rho_{\rm b}\left[\sqrt{1+\left(\frac{\sigma_{\rm m}}{d\rho_{\rm b}}\right)^2}-1\right].
\ee

In Fig.~\ref{fig2}(c), the comparison of the disks and the solid curves shows that the Donnan pressure~(\ref{e25}) provides a highly accurate approximation of the MF pressure~(\ref{net}). Thus, we use Eq.~(\ref{e25}) to provide analytical insight into the effect of the solute size on the short-range intermembrane repulsion. In this regime corresponding to a counterion-only liquid characterized by the inequalities $\sigma_{\rm m}/(\rho_{\rm b}d)\gg1$, $d\ll a$ , and $d\ll\alpha$, the expansion of Eqs.~(\ref{e25}) and~(\ref{dpb}) for short separation distances yields
\bea\label{e32}
\beta\pn&\approx&\frac{4\sigma_{\rm m}}{d}-2\rho_{\rm b}\\
&&-\frac{2\sigma_{\rm m} d}{3\alpha^2}\left\{1+\frac{\sqrt\pi a}{\alpha}e^{\frac{a^2}{\alpha^2}}\left[1+{\rm erf}\left(\frac{a}{\alpha}\right)\right]\right\};\nonumber\\
\label{e33}
\beta\pn^{({\rm PB})}&\approx&\frac{2\sigma_{\rm m}}{d}-2\rho_{\rm b}+\frac{\rho_{\rm b}^2}{\sigma_{\rm m}}d.
\eea

In Fig.~\ref{fig2}(c), Eqs.~(\ref{e32})-(\ref{e33}) are displayed by the dotted black curves. The comparison of the first terms of these asymptotic laws indicates that at short distances, finite solute size and flexibility amplify the intermembrane repulsion mediated by point charges by a factor of two.  The corresponding effect is equally illustrated in the insets of Figs.~\ref{fig2}(a)-(b). One sees that as the solute flexibility or size increases, the pressure quickly rises above the PB value and saturates at a plateau. Then, as the membrane walls get closer, this plateau approaches  from above the value of two, i.e. $P_{\rm net}/P^{\rm (PB)}_{\rm net}\to2^+$ for $\bd\ll1$.

In order to identify the origin of this effect, in Fig.~\ref{fig4}(a)-(b), we reported the positive and negative terminal charge densities of the interfacially adsorbed cations  for various $\alpha$ values. One notes that despite the opposite interactions of these terminal charges with the anionic membranes, the contact densities of the positive divalent charge and its negative monovalent counterpart dragged by the former to intermembrane region are significantly close. This feature causes in turn the nearly identical momentum exchange of the positive and negative terminal charges with the anionic membranes. As a result, the net force exerted by an extended solute molecule on the membranes is twice as strong as that induced by a point charge of the same valency. This leads to the twofold enhancement of the osmotic pressure by the solute size.

\section{SC regime of macromolecular interactions in structured liquids}

\label{sc}

In order to verify the general validity of the twofold amplification of the intermembrane repulsion by finite solute size, which has been predicted in the electrostatic MF regime of weakly charged membranes and low solute valencies, we evaluate here the intermembrane pressure in the opposite SC regime of strong membrane charges and high solute valencies. With the aim to assess the effect of the specific solute geometry, we also consider the case of the liquids composed of spherical solute molecules. 

\subsection{Review of the SC formalism}

\label{revsc}

Our SC analysis will follow the general lines of the SC formalism developed by Moreira and Netz for counterion-only liquids ~\cite{NetzPr1,NetzPr2,MoreiraCont}. The MC simulations of the aforementioned works showed that the interactions between the charged membranes and the multivalent counterions give rise to an interfacial monolayer ion distribution characterized by large lateral correlation holes. As a result, the leading order SC approximation is equivalent to the virial treatment of these dilute counterions. Thus, the SC pressure follows from the contact theorems~(\ref{pr9}) and~(\ref{s5}) via the evaluation of the field average~(\ref{fieldav})  in a pure solvent background, i.e. $\lan F[\phi(\br)]\ran\approx\lan F[\phi(\br)]\ran_0$, with
\be
\label{fieldav0}
\lan F[\phi(\br)]\ran_0=\frac{1}{Z_0}\int\frac{\mathcal{D}\phi}{\sqrt{{\rm det}\left(v\ce\right)}}\;e^{-\beta H_0[\phi]}F[\phi(\br)],
\ee
where the Hamiltonian of the solute-free liquid is
\bea
\label{h0}
\beta H_0[\phi]&=&\frac{1}{2}\int\mathrm{d}^3\br\mathrm{d}^3\br'\phi(\br)v\ce^{-1}(\br,\br')\phi(\br')\\
&&-i\int\mathrm{d}^3\br\sigma(\br)\phi(\br),\nonumber
\eea
and the corresponding partition function reads
\be
\label{scpart1}
Z_0=\int\frac{\mathcal{D}\phi}{\sqrt{{\rm det}\left(v\ce\right)}}\;e^{-\beta H_0[\phi]}.
\ee

The field averages will be calculated by splitting the potential into its average value $\phi\ce(\br)$ and the fluctuating component $\psi(\br)$ as 
\be
\label{sp1}
\phi(\br)=i\phi\ce(\br)+\psi(\br),
\ee
where the average potential satisfies the Poisson Eq.
\be
\label{sp2}
\frac{k_{\rm B}T}{e^2}\nabla\cdot\epsilon(\br)\nabla\phi\ce(\br)=-\sigma(\br).
\ee
In Appendix~\ref{red}, we show that by substituting the superposition identity~(\ref{sp1}) into Eq.~(\ref{h0}), and using the Gauss' theorem~\cite{jackson} together with the Poisson identity~(\ref{sp2}), the Hamiltonian can be simplified to
\be
\label{mt7}
\beta H_0[\psi]=\frac{k_{\rm B}T}{2e^2}\int\mathrm{d}^3\br\;\e(\br)\left[\nabla\psi(\br)\right]^2+\frac{1}{2}\int\mathrm{d}^3\br\sigma(\br)\phi\ce(\br).
\ee
Plugging Eqs.~(\ref{sp1}) and~(\ref{mt7}) into the field average in Eq.~(\ref{fieldav0}), the latter follows as a functional integral over the fluctuating potential $\psi(\br)$,
\be
\label{sp4}
\lan F[\phi(\br)]\ran_0=\int\frac{\mathcal{D}\psi}{\sqrt{{\rm det}\left(v\ce\right)}}\;e^{-\beta H'_0[\psi]}F[\psi(\br)+i\phi\ce(\br)],
\ee
where we introduced the reduced Hamiltonian
\be
\label{sp3}
\beta H'_0[\psi]=\frac{1}{2}\int\mathrm{d}^3\br\mathrm{d}^3\br'\psi(\br)v\ce^{-1}(\br,\br')\psi(\br').
\ee
In the remainder, the functional averages will be evaluated with the use of the identity~(\ref{sp4}).

For the planar charge density~(\ref{e10}), Eq.~(\ref{sp2}) solved by the potential $\phi\ce(\br)$ takes the one dimensional form
\be\label{sp2II}
\partial_z^2\phi\ce(z)=4\pi\ell_{\rm B}\sigma_{\rm m}\left[\delta(z)+\delta(z-d)\right].
\ee
According to the superposition principle, the intermembrane potential solving the linear Poisson equation~(\ref{sp2II}) is given by the sum of the individual potential components induced by the left membrane surface $\phi_{\rm L}(\br)=2\pi\ell_{\rm B}\sigma_{\rm m}z$ and the right surface $\phi_{\rm R}(\br)=2\pi\ell_{\rm B}\sigma_{\rm m}(d-z)$, i.e.
\be\label{sp5}
\phi\ce(\br)=\phi_{\rm L}(\br)+\phi_{\rm R}(\br)=2\pi\ell_{\rm B}\sigma_{\rm m} d.
\ee
Substituting now the Hamiltonian~(\ref{mt7}) into Eq.~(\ref{scpart1}), the partition function reduces to $Z_0=e^{-\int\mathrm{d}^3\br\sigma(\br)\phi\ce(\br)/2}$. Thus, together with the potential~(\ref{sp5}), the SC partition function of the interacting membranes follows in the form
\be
\label{scpart2}
Z_0=e^{2\pi\ell_{\rm B}\sigma_{\rm m}^2Sd}.
\ee

\subsection{Solute structure effects in dielectrically continuous systems}

\subsubsection{Linear solute molecules}
\label{lsc}

We consider first the SC regime of the linear solute molecules whose charge distribution is given by Eq.~(\ref{cs}). Evaluating the terminal charge densities~(\ref{e7II})-(\ref{e8II}) according to Eq.~(\ref{sp4}), and carrying-out the integrals over the intramolecular solute fluctuations, after some algebra, one gets
\be\label{sc1}
\rho_{+,p}(z)=\rho_{+,e}(z)=\lambda_+e^{-2\pi\ell_{\rm B}q_+\sigma_{\rm m} d}\gamma_+(z),
\ee
with the steric function $\gamma_+(z)$ given by Eq.~(\ref{e17}).

In order to fix the fugacity $\lambda_+$ from the electroneutrality condition
\be
\label{sc2}
S\int_0^d\mathrm{d}z\rho\ce(z)=2S\sigma_{\rm m},
\ee
one has to evaluate the solute charge density~(\ref{pr4}). By calculating the gaussian field average in Eq.~(\ref{pr4}) according to Eq.~(\ref{sp4}), accounting for the solute charge structure~(\ref{cs}), and the steric potentials~(\ref{s1}) and~(\ref{st1}), and performing the integrals over the intermembrane region and the intramolecular solute fluctuations, one obtains
\be
\label{sc3}
\rho\ce(z)=\lambda'_+q_+e^{-2\pi\ell_{\rm B}q_+\sigma_{\rm m} d}\gamma_+(z).
\ee
Substituting Eq.~(\ref{sc3}) into Eq.~(\ref{sc2}), the fugacity follows as  $\lambda_+=2\sigma_{\rm m} e^{2\pi\ell_{\rm B}q_+\sigma_{\rm m} d}/(q_+dt_+)$, where $t_+$ is the steric partition coefficient given by Eq.~(\ref{e21}). Thus, the terminal charge densities in Eq.~(\ref{sc1}) become
\be
\label{sc4}
\rho_{+,p}(z)=\rho_{+,e}(z)=\frac{2\sigma_{\rm m}}{q_+dt_+}\gamma_+(z).
\ee

Plugging Eq.~(\ref{sc4}) into the contact-value identity~(\ref{pr9}), the intermembrane pressure finally follows in the form
\be
\label{sc5}
\beta P=\frac{4\sigma_{\rm m}}{q_+d}\frac{\gamma_+(d)}{t_+}-2\pi\ell_{\rm B}\sigma_{\rm m}^2.
\ee
In the point-ion limit, the SC pressure of Refs.~\cite{NetzCont,MoreiraCont} emerges from Eq.~(\ref{sc5}) as
\be
\label{sc6}
\beta P_0=\lim_{a,\alpha\to0}\beta P=\frac{2\sigma_{\rm m}}{q_+d}-2\pi\ell_{\rm B}\sigma_{\rm m}^2.
\ee

\begin{figure}
\includegraphics[width=1.\linewidth]{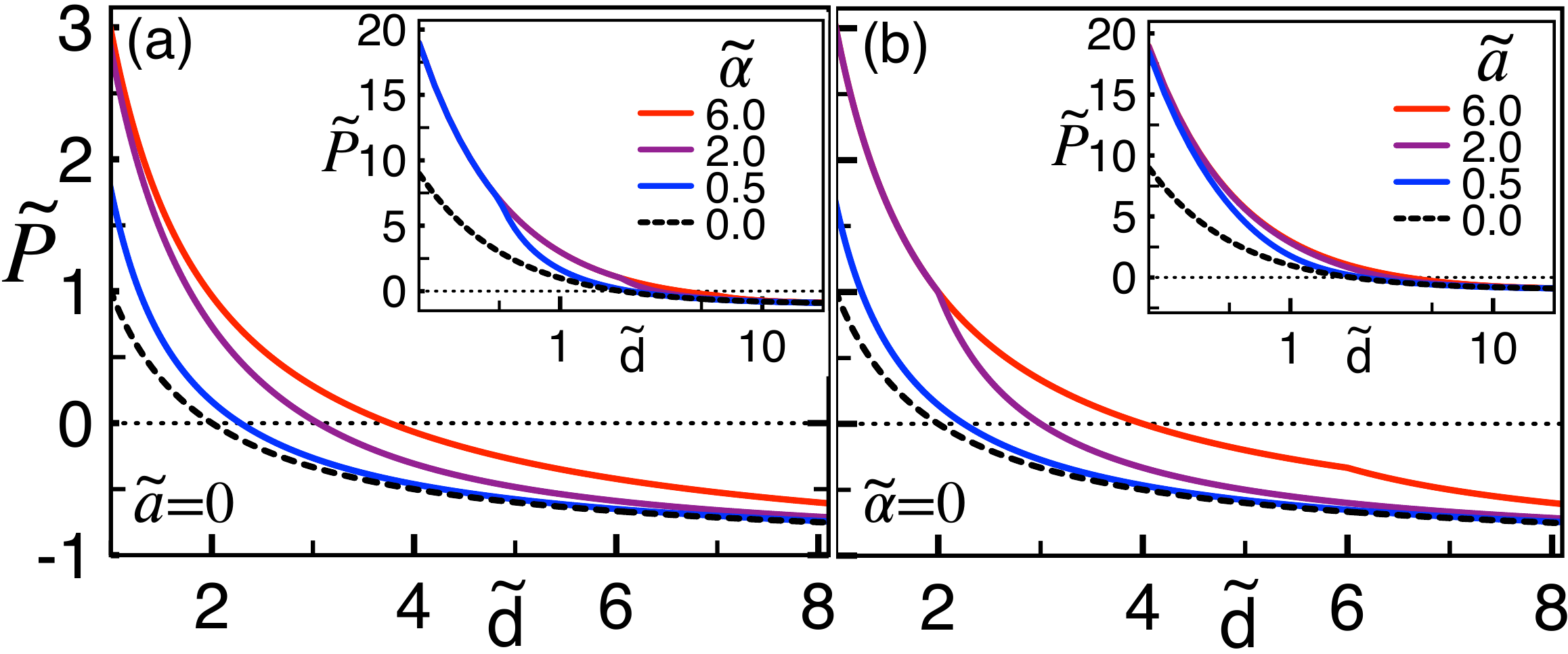}
\caption{(Color online) (a) SC-level pressure~(\ref{sc5}) renormalized as $\tilde{P}=\beta P/(2\pi\ell_{\rm B}\sigma_{\rm m}^2)$ against the rescaled distance $\tilde{d}=d/\mu$ for flexible solute molecules, and (b) its rigid solute limit~(\ref{sc8}). The semi-log plots in the insets display the pressure curves over a larger distance interval. The solute size ($\tilde{a}=a/\mu$) or flexibility ($\tilde{\alpha}=\alpha/\mu$) for each curve and symbol is indicated in the legend of its panel by the same color.}
\label{fig5}
\end{figure}

In Figs.~\ref{fig5}(a) and (b), the intermembrane pressure~(\ref{sc5}) (solid curves) is compared with its point-charge limit~(\ref{sc6}) (dashed curves) for flexible and rigid molecules, respectively. One sees that the short-range repulsive branch of the pressure increases with the molecular flexibility (size) and saturates at $\alpha\gtrsim d$ ($a\gtrsim d$). As a result, the equilibrium separation distance $d^*$ at vanishing pressure rises with the molecular extension, i.e. $\alpha\uparrow d^*\uparrow$ ($a\uparrow d^*\uparrow$). 

In accordance with the MF analysis of Sec.~\ref{res}, the comparison of the point ion pressure~(\ref{sc6}) and the rigid molecule limit of the structured solute pressure~(\ref{sc5}) 
\be
\label{sc8}
\lim_{\alpha\to0}\beta P=\frac{4\sigma_{\rm m}}{q_+d}\theta(a-d)+\frac{2\sigma_{\rm m}}{q_+(d-a/2)}\theta(d-a)-2\pi\ell_{\rm B}\sigma_{\rm m}^2
\ee
indicates that the amplification of the short-range intermembrane repulsion by the solute size corresponds to a pressure increment by a factor of two. This effect is equally illustrated in the insets of Figs.~\ref{fig5}(a)-(b). In addition, the equilibrium distance following from Eq.~(\ref{sc8})
\be
\label{eqp}
d^*=\left(2\mu+\frac{a}{2}\right)\theta(4\mu-a)+4\mu\;\theta(a-4\mu)
\ee
grows linearly with the molecular size $a$ from the point-charge value $d^*=2\mu$ to $d^*=4\mu$ where it saturates for large molecular sizes $a>4\mu$. Hence, the equilibrium distance of the point charge model is equally doubled by the solute extension.

\begin{figure}
\includegraphics[width=1.\linewidth]{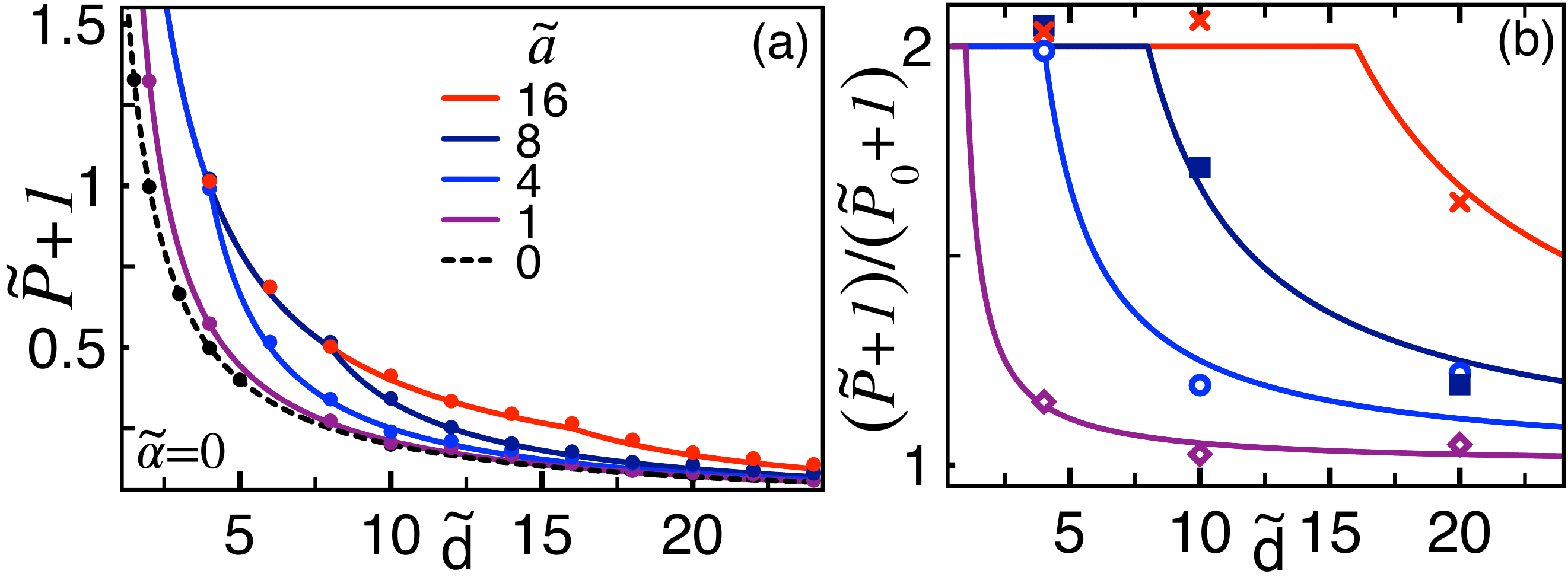}
\caption{(Color online) (a) Repulsive component of the SC pressure~(\ref{sc8}), and (b) its ratio with the point charge limit~(\ref{sc6}). The black and colored symbols are MC data from Refs.~\cite{NetzPr1} ($\Xi=10^5$) and~\cite{pincus} ($\Xi=10^4$), respectively~\cite{rem2}. The solute size for each curve and symbol is indicated in (a) by the same color.}
\label{fig6}
\end{figure}

For the sake of consistency, these conclusions have to be corroborated by numerical simulations. To the best of our knowledge, simulation data for linear solute molecules carrying opposite terminal charges is not available in the literature. However, we note that in the case of solute molecules with similar terminal charges $p_+=e_+=q$ and molecular valency $q_+=2q$, the rigid solute limit~(\ref{sc8}) of the pressure~(\ref{sc5}) is equivalent to the analytical pressure formula of Ref.~\cite{pincus}. Thus, the twofold pressure enhancement by solute size can be  probed by comparing the simulation data of Ref.~\cite{NetzPr1}  for point charges with the data of Ref.~\cite{pincus} obtained for dumbbell-like ions composed of equal terminal charges. 

In Figs.~\ref{fig6}(a)-(b), this comparison is displayed in terms of the repulsive pressure component $\beta P+2\pi\ell_{\rm B}\sigma^2_{\rm m}$  for point ions (black curve and symbols) and finite-size molecules (colored curves and symbols) together with their ratio~\cite{rem2}. In agreement with the theoretical curves, the simulation data shows that upon the increase of the molecular size at fixed intermembrane distance, or the reduction of the separation distance at fixed solute size, the repulsive force component rises monotonically  and reaches at $a\geq d$ twice the value of the point ion limit~(\ref{sc6}). Hence, the twofold pressure enhancement by solute size is equally supported by MC simulations.

\begin{figure}
\includegraphics[width=1.\linewidth]{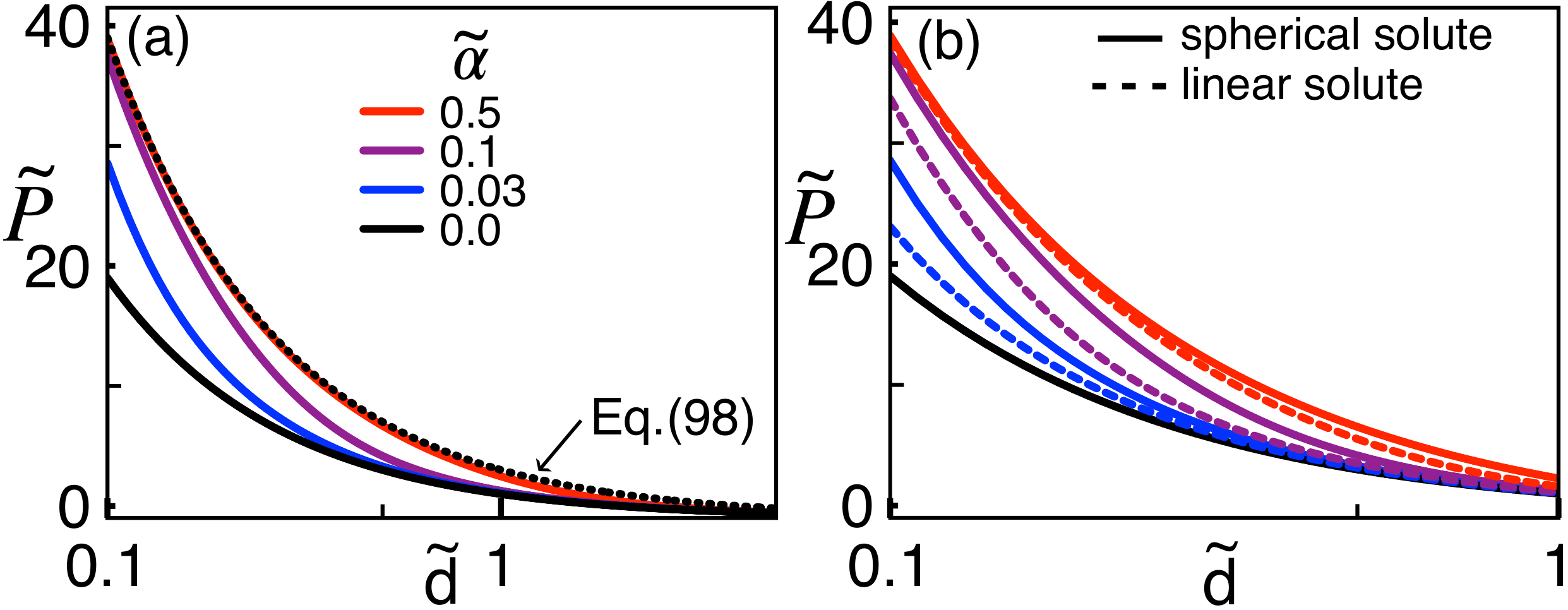}
\caption{(Color online) (a) Pressure mediated by spherical molecules (solid curves), and (b) its comparison with that mediated by linear molecules (dashed curves) for $b_0=a=0$.}
\label{fig7}
\end{figure}

\subsubsection{Spherical solute molecules}
\label{spm}

With the aim to identify the effect of the solute specificity on the intermembrane force, we evaluate here the pressure~(\ref{s5}) in solutions composed of the spherical solute counterions depicted in Fig.~\ref{fig1}(c). Calculating the field-average in Eq.~(\ref{s5}) according to Eq.~(\ref{sp4}), and carrying-out the integral over the radial solute size, the intermembrane pressure follow as
\bea
\label{sc9II}
\hspace{-5mm}&&\beta P=\lambda_+e^{-2\pi\ell_{\rm B}q_+\sigma_{\rm m} d}
\frac{{\rm erf}\left[\frac{d}{2\alpha}\right]-{\rm erf}\left(\frac{b_0}{\alpha}\right)}{1-{\rm erf}\left(\frac{b_0}{\alpha}\right)}-2\pi\ell_{\rm B}\sigma_{\rm m}^2.\nonumber\\
\hspace{-5mm}
\eea

The fugacity in Eq.~(\ref{sc9II}) will be determined again from the pore electroneutrality condition~(\ref{sc2}). First, by carrying out the field average in the solute charge density~(\ref{pr4}) according to Eq.~(\ref{sp4}), accounting for the solute structure function in Eq.~(\ref{cs2}), the one dimensional Drude potential in Eq.~(\ref{dr2}), and the steric constraint in Eq.~(\ref{s2}), and plugging the resulting expression for the solute charge density into Eq.~(\ref{sc2}), the solute fugacity follows as
\be
\label{sc10}
\lambda_+=\frac{2\sigma_{\rm m}}{G(d)d}e^{2\pi\ell_{\rm B}q_+\sigma_{\rm m} d},
\ee
where we defined the coefficient
\bea
\label{sc11}
G(d)&=&\int_0^d\frac{\mathrm{d}z}{d}\int\mathrm{d}^2\br'_\pa\int_{b_0}^{d-b_0}\mathrm{d}z'\\
&&\hspace{0cm}\times\int_{b_0}^{b_+(z')}\frac{\mathrm{d}b}{\sqrt{\pi}\alpha c_i}\;e^{-\frac{b^2}{\alpha^2}}\hat{n}_+\left(||\br'-\br||,b\right).\nonumber
\eea

The integrals in Eq.~(\ref{sc11}) have been evaluated with an isotropic surface charge distribution of the form
\be
\label{sc12}
\hat{n}_+\left(||\br'-\br||,b\right)=n_{\rm s}\left(\Delta r'-b\right)=\frac{q_+}{4\pi b^2}\delta\left(\Delta r-b\right),
\ee
with the shortcut notation $\Delta r=||\br'-\br||$. The corresponding calculation reported in Appendix~\ref{X} yields
\bea
\label{sc14}
G(d)&=&\frac{1}{2c_i}\left\{{\rm erf}\left(\frac{d}{2\alpha}\right)-{\rm erf}\left(\frac{b_0}{\alpha}\right)\right.\\
&&\hspace{9mm}\left.-\frac{2\alpha}{\sqrt\pi d}\left[e^{-\frac{b_0^2}{\alpha^2}}-e^{-\frac{d^2}{4\alpha^2}}\right]\right\}.\nonumber
\eea
Hence, substituting the fugacity~(\ref{sc10}) into Eq.~(\ref{sc9II}), the intermembrane pressure finally becomes
\be
\label{sc16}
\beta P=\frac{2\sigma_{\rm m}}{q_+T(d)d}-2\pi\ell_{\rm B}\sigma_{\rm m}^2,
\ee
with the auxiliary function
\be
\label{sc17}
T(d)=1-\frac{2\alpha}{\sqrt\pi d}\frac{e^{-\frac{b_0^2}{\alpha^2}}-e^{-\frac{d^2}{4\alpha^2}}}{{\rm erf}\left(\frac{d}{2\alpha}\right)-{\rm erf}\left(\frac{b_0}{\alpha}\right)}.
\ee

In the limit of purely rigid solute molecules of radius $b_0$, the pressure~(\ref{sc16}) simplifies to
\be
\label{sc18}
\lim_{\alpha\to0}\beta P=\frac{2\sigma_{\rm m}}{q_+(d-2b_0)}-2\pi\ell_{\rm B}\sigma_{\rm m}^2.
\ee
Thus, at the SC level, the spherical hard-core size simply leads to the renormalization of the intermembrane distance in Eq.~(\ref{sc6}) by the molecular diameter. Then, at short separation distances, the pressure~(\ref{sc16}) becomes
\be\label{sc19}
\beta P\approx\frac{4\sigma_{\rm m}}{q_+(d-2b_0)}-2\pi\ell_{\rm B}\sigma_{\rm m}^2.
\ee
Fig.~\ref{fig7}(a) shows that as the flexibility of the spherical molecules increases, the interaction pressure~(\ref{sc16}) rises monotonically from the rigid solute limit~(\ref{sc18}) towards the upper value~(\ref{sc19}) where it saturates at $\alpha\approx d$. 

Hence, the size-induced twofold enhancement of the short range pressure with linear solute molecules is equally caused by a radially fluctuating spherical charge spread. Finally, Fig.~\ref{fig7}(b) shows that the pressure enhancement by the spherical fluctuations is generally stronger than that caused by the linear fluctuations. However, as the molecular extension becomes comparable with the intermembrane distance, the solute specificity is quickly lost, and the pressure mediated by both types of molecules tends to the upper limit in Eq.~(\ref{sc19}).

\begin{figure*}
\includegraphics[width=1.\linewidth]{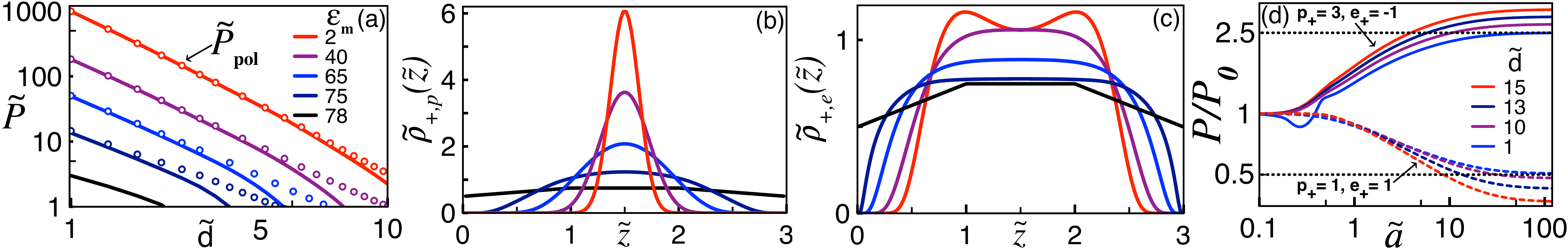}
\caption{(Color online) (a) Pressure~(\ref{g12}) (solid curves) and its polarization component~(\ref{g15}) (circles). (b)-(c) Terminal charge densities~(\ref{g11}) renormalized as $\tilde{\rho}_{+,j}(\tilde{z})=\rho_{+,j}(\tilde{z})/(2\pi\ell_{\rm B}\sigma_{\rm m}^2)$  at the separation distance $\tilde{d}=3$. The membrane permittivity of each curve is indicated in (a). The divalent counterions ($q_+=2$) with size $\tilde{a}=2$ have terminal charges $p_+=3$ and $e_-=-1$.  (d) Ratio of the pressure~(\ref{g12}) and its point-charge limit~(\ref{g16}) against the solute size at various membrane distances for opposite (solid curves) and similar terminal charges (dashed curves) at the permittivity $\e_{\rm m}=2$. The coupling parameter is $\Xi=100$. }
\label{fig8}
\end{figure*}

\subsection{Intramolecular structure effects in dielectrically discontinuous systems}

In this part, we probe the solute structure effects on the intermembrane interactions by taking into account the dielectric contrast between the electrolyte and the insulating membrane. As the contact theorem~(\ref{pr9}) does not account for the resulting polarization forces, we evaluate the intermembrane interactions mediated by linear solute molecules directly from the liquid grand potential.

\subsubsection{Evaluation of the pressure from the grand potential}

The SC-level grand potential will be derived from the dilute counterion approximation explained in Sec.~\ref{revsc}. Following these lines, the virial expansion of the liquid grand potential $\beta\Omega_{\rm G}=-\ln Z_{\rm G}$ for the partition function~(\ref{par}) and the Hamiltonian~(\ref{e5II}) yields
\bea
\label{g1}
\beta\Omega_{\rm G}&\approx&-\ln Z_0\\
&&-\frac{\lambda_+}{c_+}\int\frac{\mathrm{d}^3\bb}{(\sqrt{\pi}\alpha)^3}e^{-\frac{(b-a)^2}{\alpha^2}}\nonumber\\
&&\hspace{2mm}\times\int\mathrm{d}^3\br\;e^{\epsilon_+-w_i(\br,\bb)}\lan e^{i\left[p_+\phi(\br)+e_+\phi(\br+\bb)\right]}\ran_0,\nonumber
\eea
with the solute-free partition function $Z_0$ given by~(\ref{scpart2}), and the field average defined by Eq.~(\ref{fieldav0}). 

In order to simplify the numerical evaluation of the interaction pressure, we will consider the rigid solute limit where the Drude distribution function becomes
\be
\label{g2}
\lim_{\alpha\to0}\frac{e^{-\frac{(b-a)^2}{\alpha^2}}}{c_+\left(\sqrt{\pi}\alpha\right)^3}=\frac{1}{4\pi a^2}\delta(b-a).
\ee
Applying the identity~(\ref{g2}) to the SC-expanded grand potential~(\ref{g1}), substituting the potential decomposition~(\ref{sp1}) to evaluate the field average with Eq.~(\ref{sp4}), and incorporating the steric potential in Eqs.~(\ref{s1}) and~(\ref{st1}) together with the solute self-energy~(\ref{se}), one obtains
\bea
\label{g3}
\beta\Omega_{\rm G}&=&-2\pi\ell_{\rm B}\sigma_{\rm m}^2dS\\
&&-S\lambda_+e^{-2\pi\ell_{\rm B}q_+\sigma_{\rm m}d}\int_0^d\mathrm{d}z\int_{a_-(z)}^{a_+(z)}\frac{\mathrm{d}a_z}{2a}\;e^{-U_p(\ba,z)},\nonumber
\eea
where we introduced the auxiliary functions 
\be
\label{g4}
a_-(z)=-{\rm min}(a,z);\hspace{5mm}a_+(z)={\rm min}(a,d-z).
\ee
In Eq.~(\ref{g3}), the dielectric potential experienced by the C.M. charge $p_+$ reads
\bea
\label{g5}
U_p(\ba,z)&=&\frac{1}{2}\left\{p_+^2\delta v\ce(0,z,z)+e_+^2\delta v\ce(0,z+b_z,z+b_z)\right.\nonumber\\
&&\hspace{5mm}+\left.2p_+e_+\delta v\ce(\ba_\pa,z,z+b_z)\right\},
\eea
with the renormalized Green's function
\be
\label{g6}
\delta v\ce(\br_\pa-\br'_\pa,z,z')=v\ce(\br,\br')-v_{\rm c,b}(\br-\br').
\ee

Next, we evaluate the counterion fugacity from the electroneutrality constraint~(\ref{sc2}). To this aim, we first calculate the terminal charge densities~(\ref{e7II})-(\ref{e8II}) via the potential decomposition in Eq.~(\ref{sp1}) and the field average defined by Eq.~(\ref{sp4}). Accounting again for the solute self-energy~(\ref{se}) and the steric potential~(\ref{s1}), and taking the rigid solute limit~(\ref{g2}), one obtains
\be\label{g7}
\rho_{+,j}(z)=\lambda_+e^{-2\pi\ell_{\rm B}q_+\sigma_{\rm m} d}\int_{a_-(z)}^{a_+(z)}\frac{\mathrm{d}a_z}{2a}e^{-U_j(\ba,z)},
\ee
where we introduced the index $j=\{p,e\}$, and the dielectric potential experienced by the terminal charge $e_i$,
\bea
\label{g8}
U_e(\ba,z)&=&\frac{1}{2}\left\{p_+^2\delta v\ce(0,z+b_z,z+b_z)+e_+^2\delta v\ce(0,z,z)\right.\nonumber\\
&&\hspace{5mm}+\left.2p_+e_+\delta v\ce(\ba_\pa,z,z+b_z)\right\}.
\eea
Thus, plugging the solute charge density $\rho\ce(z)=p_+\rho_{+,p}(z)+e_+\rho_{+,e}(z)$ into the electroneutrality identity~(\ref{sc2}), the counterion fugacity follows as
\be
\label{g9}
\lambda_+=\frac{2\sigma_{\rm m}}{J_+d}e^{2\pi\ell_{\rm B}q_+\sigma_{\rm m} d},
\ee
with the auxiliary coefficient
\be
\label{g10}
J_+=\int_0^d\frac{\mathrm{d}z}{d}\int_{a_-(z)}^{a_+(z)}\frac{\mathrm{d}a_z}{2a}\left\{p_+e^{-U_p(\ba,z)}+e_+e^{-U_e(\ba,z)}\right\}.
\ee
Finally, substituting the fugacity~(\ref{g9}) into Eq.~(\ref{g7}), the terminal charge densities become for $j=\{p,e\}$
\be\label{g11}
\rho_{+,j}(z)=\frac{2\sigma_{\rm m}}{J_+d}\int_{a_-(z)}^{a_+(z)}\frac{\mathrm{d}a_z}{2a}e^{-U_j(\ba,z)}.
\ee
The functional form of the self-energies $U_{\{p,e\}}(\ba,z)$ in Eqs.~(\ref{g5}) and~(\ref{g8}) are reported in Appendix~\ref{ter}.

Evaluating now the variation of the grand potential~(\ref{g3}) with respect to the intermembrane distance according to Eq.~(\ref{pr1}), and replacing the fugacity by Eq.~(\ref{g9}), after some lengthy but straightforward algebra, the SC-level interaction pressure finally follows as
\be
\label{g12}
P=P_{\rm el}+P_{\rm ent}+P_{\rm pol}.
\ee
In Eq.~(\ref{g12}), the attractive electrostatic pressure is
\be
\label{g13}
\beta P_{\rm el}=-2\pi\ell_{\rm B}\sigma_{\rm m}^2\left\{1+\frac{e_+}{\sigma_{\rm m}}\int_0^d\mathrm{d}z\left[\rho_{+,p}(z)-\rho_{+,e}(z)\right]\right\}.
\ee
Then, the repulsive osmotic pressure induced by the solute-membrane collisions reads
\be\label{g14}
\beta P_{\rm ent}=\rho_{+,p}(d)+\rho_{+,e}(d).
\ee
Finally, the repulsive pressure component associated with the polarization forces is
\be
\label{g15}
\beta P_{\rm pol}=-\frac{2\sigma_{\rm m}}{J_+}\int_0^d\frac{\mathrm{d}z}{d}\int_{a_-(z)}^{a_+(z)}\frac{\mathrm{d}a_z}{2a}\;e^{-U_p(\ba,z)}\frac{\partial U_p(\ba,z)}{\partial d}.
\ee

In the point charge limit $a\to0$, the interaction force~(\ref{g12}) reduces to the limit of infinitely thick membranes of the point charge pressure derived in Ref.~\cite{KanducSC2},
\bea
\label{g16}
\beta P_0=\lim_{\alpha\to0}\lim_{a\to0}\beta P&=&-2\pi\ell_{\rm B}\sigma_{\rm m}^2+\frac{2\sigma_{\rm m}}{q_+I_0d}e^{-U_0(d)}\\
&&-\frac{2\sigma_{\rm m}}{q_+I_0}\int_0^d\frac{\mathrm{d}z}{d}e^{-U_0(z)}\frac{\partial U_0(z)}{\partial d},\nonumber
\eea
where $I_0=\int_0^d\mathrm{d}z\;e^{-U_0(z)}/d$. The functional form of the image-charge potential $U_0(z)$ is reported in Appendix~\ref{ter}.

\subsubsection{Polarization effects on intermembrane interactions}

We first consider the case of the solute molecules with opposite terminal charges. This geometry qualitatively accounts for the charge spread of ionized atoms and zwitterionic molecules. Fig.~{\ref{fig8}(a) displays the pressure~(\ref{g12}) (solid curves) mediated by divalent counterions ($q_+=2$) with terminal charges $p_+=3$ and $e_-=-1$, and size $\tilde{a}=2$. Similar to the case of point ions~\cite{KanducSC2}, the reduction of the membrane permittivity amplifies the repulsive pressure by orders of magnitude. The comparison of the curves and circles indicates that the enhanced repulsion is driven by the amplification of the polarization pressure~(\ref{g15}). This force originates from the energetic cost of inserting between the membranes a solute molecule repelled by its electrostatic images located on the membrane sides. As the intermembrane confinement enhances this cost, the counterion presence forced by electroneutrality causes the repulsion of the adjacent membranes. 

For the parameters considered here, we numerically found that the integral in Eq.~(\ref{g13}) is vanishingly small. Thus, despite the presence of the image forces, the electrostatic pressure is unaffected by the solute structure, 
\be\label{g16II}
\beta P_{\rm el}\approx-2\pi\ell_{\rm B}\sigma_{\rm m}^2.
\ee 
Moreover, in Figs.~{\ref{fig8}(b)-(c), the density plots of the terminal charges show that the repulsive image-charge interactions emerging at $\e_{\rm m}<\e_{\rm w}$ leads to a total interfacial solute exclusion, resulting in a vanishing contact density and the solute accumulation in the mid-pore region. Thus, for $\e_{\rm m}<\e_{\rm w}$, the entropic pressure component~(\ref{g14}) vanishes, $P_{\rm ent}=0$. Hence, while the repulsive branch of the pressure~(\ref{sc8}) in dielectrically continuous systems (black curve in Fig.~{\ref{fig8}(a)) stems from the solute entropy excess, the repulsive force between low permittivity membranes (colored curves and circles) originates from the polarization force~(\ref{g15}) opposed at large distances by the direct solute-membrane attraction~(\ref{g16II}).

\begin{figure}
\includegraphics[width=1.\linewidth]{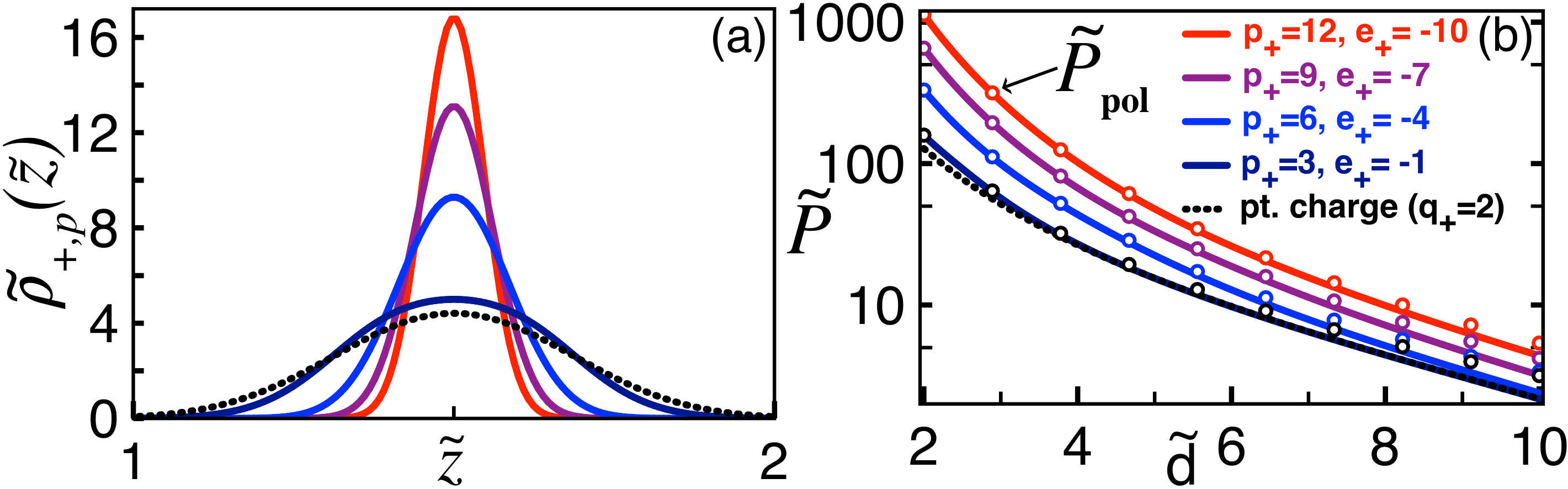}
\caption{(Color online) (a) Solute density~(\ref{g11}) at the separation distance $\tilde{d}=3$. (b) Pressure~(\ref{g12}) (curves) and its polarization component~(\ref{g15}) (circles). The terminal charges of the divalent counterions ($q_+=2$) are displayed in the legend of (b).  The coupling parameter, the membrane permittivity, and the solute size are $\Xi=100$, $\e_{\rm m}=2$, and $\tilde{a}=1$. }
\label{fig9}
\end{figure}

In Fig.~\ref{fig9}, we illustrate the alteration of the solute partition and pressure with the number of terminal charges at fixed solute valency. The plot shows that the amplification of the image-charge interactions by the rise of the terminal charges strengthens the interfacial solute depletion, sharply rises the solute density in the mid-slit, and enhances the polarization force and pressure by several factors. At the terminal charge numbers corresponding to the nuclear and electronic charges of ${\rm Mg}^{2+}$ cations (red curve), the pressure enhancement by the intramolecular charge spread is substantial even at the membrane distances exceeding the solute size by an order of magnitude.

In Figs.~{\ref{fig10}(a)-(b), we display explicitly the variation of the solute partition and pressure with the solute length. One sees that as the latter rises from the point charge limit, the increasingly strong individual coupling of the terminal charges to their electrostatic images enhances the interfacial solute depletion, and amplifies the polarization force and pressure, i.e. $a\uparrow P_{\rm pol}\uparrow P\uparrow$. Beyond the characteristic size $\tilde{a}\sim10$ where the terminal charges behave as decoupled ions, the pressure saturates at more than twice the magnitude of the point charge pressure.

Interestingly, Figs.~{\ref{fig10}(c)-(d) show that as the solutes molecules with opposite elementary charges are replaced by divalent counterions with equal terminal charges such as the putrescine molecules,  the finite size effects are reversed. Namely, the rise of the solute size weakens the interfacial solute depletion, and attenuates the net pressure together with its polarization component, i.e. $a\uparrow P_{\rm pol}\downarrow P\downarrow$. In the regime $\tilde{a}\gtrsim10$, the intermembrane force saturates at half of the point-charge pressure.

In order to elucidate the opposing effects of the solute size on the interaction of the membranes confining molecules carrying similar and opposite elementary charges, we consider the limit $a\to\infty$ where the decoupling of the intramolecular charges results in the saturation of the pressure. In the corresponding limit where the bounds of the inner integral in Eq.~(\ref{g15}) yield the constraints $a_z\to0$ and $a_\pa=\sqrt{a^2-a_z^2}\to\infty$, Eq.~(\ref{g15}) approaches the point charge limit of the polarization pressure (the third term in Eq.~(\ref{g16})) induced by the individual charges $p_i$ and $e_i$. Hence, for large solute sizes, the ratio of the intermembrane forces mediated by the linear and point-like solute molecules can be approximated by
\be
\label{g17}
\lim_{a\to\infty}\frac{P}{P_0}\approx\frac{p_+^2+e_+^2}{q_+^2}=\frac{p_+^2+e_+^2}{\left(p_++e_+\right)^2}.
\ee
For $e_+>0$ ($e_+<0$), the ratio~(\ref{g17}) is less (greater) than one. Thus, in accordance with Figs.~\ref{fig10}(b)-(d), this identity predicts the weakening (enhancement) of the intermembrane pressure by the charge spread of counterions composed of similar (opposite) terminal charges.

This prediction is directly verified in Fig.~\ref{fig8}(d) displaying the ratio $P/P_0$ against the solute length (colored curves) together with the asymptotic limit~(\ref{g17}) (horizontal lines). One sees that increasing the length of the counterions carrying similar (opposite) terminal charges, the pressure drops (rises) and saturates at a value below (above) the limit~(\ref{g17}). Then, as the membrane distance is reduced at fixed solute size $\tilde{a}\gtrsim10$, the saturation value converges towards the asymptotic limit in Eq.~(\ref{g17}).

\begin{figure}
\includegraphics[width=1.\linewidth]{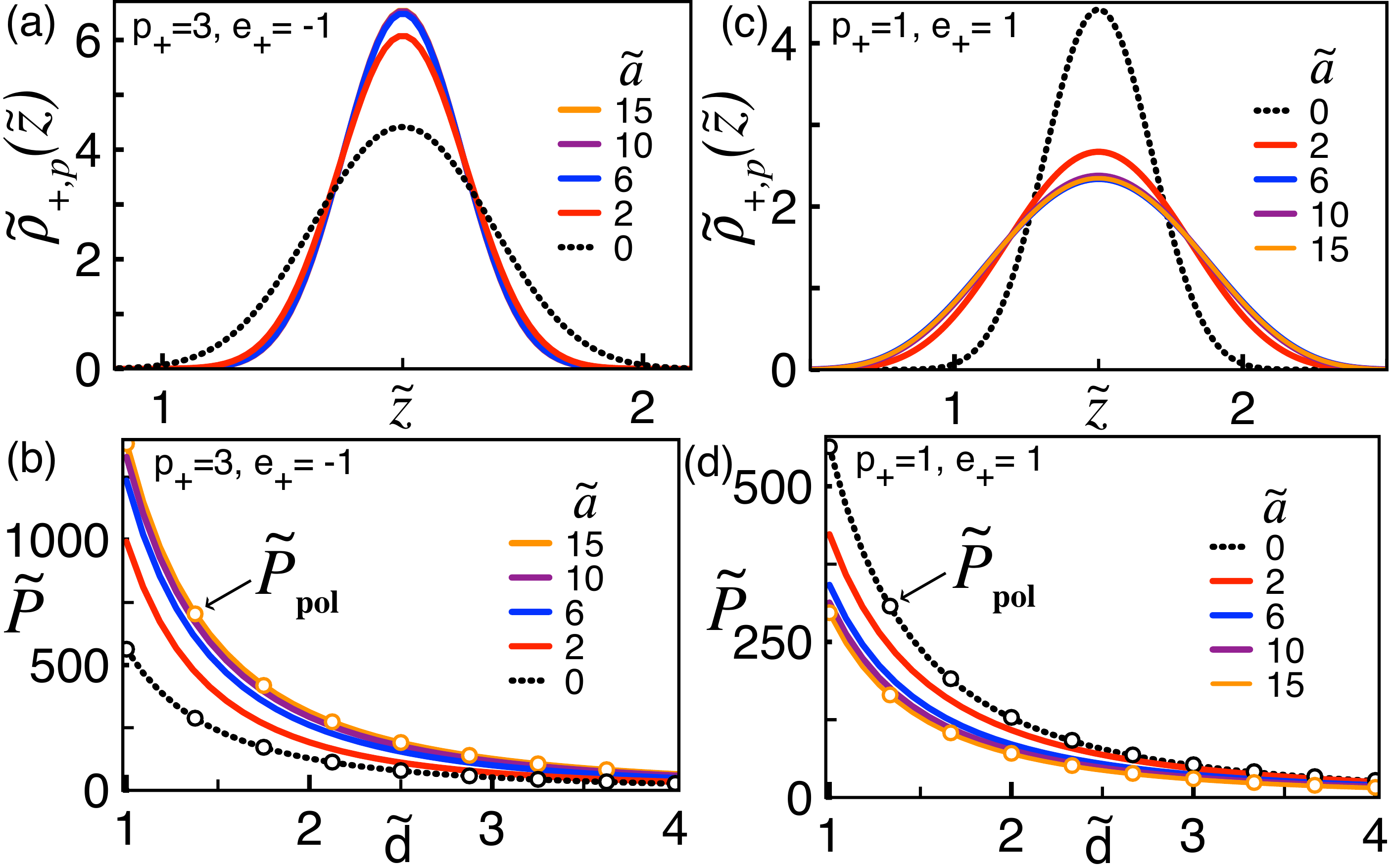}
\caption{(Color online) (a) Solute density~(\ref{g11}) at the membrane distance $\tilde{d}=3$. (b) Pressure~(\ref{g12}) (curves) and its polarization component~(\ref{g15}) (circles). The charges of the divalent counterions ($q_+=2$) are $p_+=3$ and $e_-=-1$. (c) and (d) display the curves in (a) and (b) for divalent counterions ($q_+=2$) with equal terminal charges $p_+=e_-=1$. The remaining parameters are the same as in Fig.~\ref{fig9}.}
\label{fig10}
\end{figure}

\section{Conclusions}

The complex internal structure of the solute molecules omnipresent in living systems requires the characterization of the macromolecular interactions governing biological processes beyond the point-ion approximation. In this article, we developed a field theory of intermolecular interactions explicitly incorporating the intramolecular charge structure and conformations of solute molecules. From the field-theoretic partition function of the model, we derived a generalized contact value theorem valid for any electrostatic coupling strength and solute charge composition. Via this extended contact-value identity, we characterized the effect of the solute charge spread and geometry on the electrostatic stability of like-charged membranes in the electrostatic MF and SC regimes.

In dielectrically continuous systems, the MF limit of our formalism indicates that the extended charge structure of the linear solute molecules induces a long-range depletion attraction, and the twofold enhancement of the short-range intermembrane repulsion caused by the separate momentum exchange of the terminal solute charges with the membrane walls. We found that the twofold enhancement of the short-range pressure is equally present in the opposite SC regime of linear as well as fluctuating spherical solute molecules.  Moreover, via the comparison of the MC data for point charges~\cite{NetzPr1} and dumbbell  ions~\cite{pincus} from the literature, we showed that this feature is also supported by simulations. These findings indicate the universality of this finite size effect in terms of the electrostatic coupling strength and the solute geometry.

Accounting as well for the dielectric contrast between the confined liquid and the low permittivity membranes, we showed that the resulting polarization forces significantly enhance the solute specificity of the intermembrane interactions. Indeed, in contrast with the dielectrically continuous systems where the membrane repulsion is set by the entropically driven osmotic pressure, the repulsive coupling of low permittivity membranes originates from the electrostatic cost to place between them a solute molecule repelled by its electrostatic images. The extended structure of the solute molecules carrying opposite elementary charges enhances this cost and amplifies the intermembrane force, whereas the finite size of the solute molecules bearing similar terminal charges reduces the polarization force and weakens the pressure. It was also shown that the intramolecular solute composition can significantly affect the interaction pressure even at separation distances exceeding the characteristic solute size by an order of magnitude. Thus, the solute structure effects on the coupling of low permittivity membranes is longer-ranged than that of the dielectrically continuous systems where these effects were shown to arise at distances comparable with the solute size.

We note that our model is based on an implicit solvent approach neglecting the electrostatics of water molecules. Indeed, in our recent work on explicit solvent liquids including point-like salt ions, we showed that the cancellation of various solvent effects of large magnitude but opposite sign leads to a nearly vanishing contribution from explicit solvent to intermolecular forces~\cite{PRE2022}. The explicit inclusion of solvent molecules into the present formalism will be necessary to verify the validity of this cancellation mechanism in the more general case of ions with inner charge structure.

\smallskip
\appendix

\section{Derivation of the MF-level contact value identity from the MF grand potential}
\label{cont}

In this appendix, we calculate the MF limit of the contact value identity~(\ref{pr9}) derived for linear solute molecules directly from the MF grand potential in Eq.~(\ref{e13}). The MF pressure follows from the variation of this grand potential with respect to the intermembrane distance $d$, i.e. $\beta P=-\delta f/\delta d$. This yields
\bea\label{e14}
\beta P&=&\sigma_{\rm m}\phi'(d)+\sum_{i=1}^s\rho_{i,p}(d)+\sum_{i=1}^sJ_i\\
&&-\int_{-\infty}^{\infty}\mathrm{d}z\frac{\delta f}{\delta\phi(z)}\frac{\delta\phi(z)}{\delta d},\nonumber
\eea
where we defined the integral $J_i=\int_0^d\mathrm{d}z\partial_d\rho_{i,p}(z)$. First, plugging into this integral the cationic charge density in Eq.~(\ref{e12}), using the Leibniz integral rule, and introducing the change of variable $z\to d-\bz$, one finds that the integral equals the contact density of the terminal charge $e$, i.e. $J_i=\rho_{i,e}(d)$. Then, we note that as the average potential satisfies the SPB Eq.~(\ref{e6}) obtained from the minimization of the grand potential density~(\ref{e13}), the fourth term on the r.h.s. of Eq.~(\ref{e14}) vanishes. Finally, owing to the discontinuity of the electrostatic field at the charged membrane interface, one has
\be\label{e14II}
\phi'(d)=\frac{1}{2}\left[\phi'(d^+)+\phi'(d^-)\right]=-2\pi\ell_{\rm B}\sigma_{\rm m},
\ee
where we accounted for the vanishing electric field in the ion-free membrane, i.e. $\phi'(d^+)=0$. Taking into account these simplifications in Eq.~(\ref{e14}), and subtracting the bulk osmotic pressure $\beta P_{\rm b}=\sum_{i=1}^s\rho_{i{\rm b}}$ acting on the outer surface of the interacting membranes, the net intermembrane pressure $\pn=P-P_{\rm b}$ follows as,
\be
\label{e15}
\beta\pn=\sum_{i=1}^s\left[\rho_{i,p}(d)+\rho_{i,e}(d)-\rho_{i{\rm b}}\right]-2\pi\ell_{\rm B}\sigma_{\rm m}^2,
\ee
where the MF densities are given by Eqs.~(\ref{e12})-(\ref{e12II}).

\section{Numerical solution of the SPB Eq.~(\ref{e11})}
\label{nums}

Here, we develop a recursive algorithm for the solution of the integro-differential SPB Eq.~(\ref{e11}),
\bea
\label{e11A}
&&\phi''(z)+4\pi\ell_{\rm B}\sum_{i=1}^s\left[p_i\rho_{i,p}(z)-e_i\rho_{i,e}(z)\right]\\
&&=4\pi\ell_{\rm B}\sigma_{\rm m}\left[\delta(z)+\delta(d-z)\right].\nonumber
\eea

\subsection{Monopolar limit of the SPB Eq.~(\ref{e11A})}
\label{mon}

In this part, we derive the monopolar limit of the SPB Eq.~(\ref{e11A}) whose solution will be used as the input potential for the recursive solution of this equation. The corresponding monopolar approximation neglects the electrostatic multipole moments of the linear solute molecules and accounts only for the steric exclusion effects originating from their finite size.

First, we carry out the multipolar expansion of the potential functions in the exponentials of Eq.~(\ref{e12})-(\ref{e12II}) in terms of the dipolar length $\bz$. At the monopolar order $O\left(b_z^0\right)$, the charge densities become
\be\label{e16}
\rho_{i,\alpha}(z)\approx\rho_{i{\rm b}}\gamma_i(z)e^{-q_i\phi_0(z)}
\ee
for $\alpha=\{p,e\}$, with the monopolar potential $\phi_0(z)$, and the steric exclusion function $\gamma_i(z)=\int_{-z}^{d-z}\mathrm{d}b_zg_i(b_z)$, or
\bea\label{in1A}
&&\hspace{-3mm}\gamma_i(z)=\frac{1}{c_i}\left\{\left(1+\frac{2a_i^2}{\alpha_i^2}\right){\rm erf}\left(\frac{a_i}{\alpha_i}\right)+\frac{a_id}{\alpha_i^2}\right.\\
&&\hspace{13mm}+\left[\frac{1}{2}-\frac{a_i^2}{\alpha_i^2}\left(\frac{z}{a_i}-1\right)\right]{\rm erf}\left(\frac{z-a_i}{\alpha_i}\right)\nonumber\\
&&\hspace{13mm}+\left[\frac{1}{2}-\frac{a^2_i}{\alpha_i^2}\left(\frac{d-z}{a_i}-1\right)\right]{\rm erf}\left(\frac{d-z-a_i}{\alpha_i}\right)\nonumber\\
&&\hspace{13mm}\left.-\frac{a_i}{\sqrt{\pi}\alpha_i}\left[e^{-\frac{(d-z-a_i)^2}{\alpha_i^2}}+e^{-\frac{(z-a_i)^2}{\alpha_i^2}}-2e^{-\frac{a_i^2}{\alpha_i^2}}\right]\right\}\nonumber.
\eea

Plugging the ion densities in Eq.~(\ref{e16}) into Eq.~(\ref{e11A}), the monopolar limit of the SPB equation emerges as a generalized PB equation with a non-uniform screening parameter,
\be
\label{e18}
\phm''(z)+\sum_{i=1}^s\kappa_i^2q_i\gamma_i(z)e^{-q_i\phm(z)}=4\pi\ell_{\rm B}\sigma_{\rm m}\left[\delta(z)+\delta(d-z)\right],
\ee
with the DH screening parameter per species 
\be\label{dh}
\kappa_i^2=4\pi\ell_{\rm B}\rho_{i{\rm b}}. 
\ee
The monopolar potential profile $\phm(z)$ satisfying Eq.~(\ref{e18}) can be numerically obtained via standard algorithms such as the fourth order Runge-Kutta method.

\subsection{Recursive solution of the SPB Eq.~(\ref{e11A})}
\label{rec}

In this part, we explain the recursive solution of the SPB Eq.~(\ref{e11A}) that uses the numerical solution of Eq.~(\ref{e18}). First, we recast Eq.~(\ref{e11A}) such that its l.h.s. has the form of Eq.~(\ref{e18}),
\be
\label{e18I}
\phi''(z)+\sum_{i=1}^s\kappa_i^2q_i\gamma_i(z)e^{-q_i\phi(z)}=W[\phi(z);z].
\ee
In Eq.~(\ref{e18I}), we omitted the BCs on the r.h.s. of Eq.~(\ref{e18}), and introduced the corrective source function
\bea
\label{e19}
W[\phi(z);z]&=&\sum_{i=1}^s\kappa_i^2\left\{q_i\gamma_i(z)e^{-q_i\phi(z)}\right.\\
&&\hspace{1.2cm}\left.-p_ik_{i,p}(z)+e_ik_{i,e}(z)\right\}.\nonumber
\eea
The charge partition functions in Eq.~(\ref{e19}) are related to the ion densities in Eq.~(\ref{e16}) as $k_{i,p}(z)=\rho_{i,p}(z)/\rho_{i,{\rm b}}$ and $k_{i,e}(z)=\rho_{i,e}(z)/\rho_{i,{\rm b}}$.

In this work, Eq.~(\ref{e18I}) was solved iteratively via a standard Runge-Kutta algorithm. This iterative solution is based on the treatment of the source function $W[\phi(z);z]$ as an external function. This has been achieved by including in Eq.~(\ref{e18I}) the corresponding source function evaluated with the potential profile of the preceding iterative step. To this aim, we modify Eq.~(\ref{e18I}) as
\be
\label{e18II}
\phi''_j(z)+\sum_{i=1}^s\kappa_i^2q_i\gamma_i(z)e^{-q_i\phi_j(z)}=W[\phi_{\alpha-1}(z);z],
\ee
where the index $j$ denotes the number of the iterative step. The recursive cycle was initialized at $\alpha=1$ by solving first Eq.~(\ref{e18}) to obtain the potential profile $\phi_0(z)$. Then, this monopolar potential was used to evaluate the source term of Eq.~(\ref{e18II}), and to obtain from the Runge-Kutta algorithm the updated potential $\phi_1(z)$. At the next iterative step $\alpha=2$, Eq.~(\ref{e18II}) was solved by using the updated potential $\phi_1(z)$ as the new input potential in the source function on the r.h.s., and this cycle was continued until numerical convergence is achieved. For the model parameters used in this work, the corresponding scheme enabled the rapid convergence of the solution cycle in less than five iterative steps.

\section{Derivation of the Hamiltonian~(\ref{mt7})}
\label{red}

In this appendix, we explain the derivation of the reduced Hamiltonian~(\ref{mt7}). First, by plugging the decomposition~(\ref{sp1}) into the SC Hamiltonian~(\ref{h0}), one obtains
\bea
\label{h1}
\beta H_0[\psi]&=&\frac{k_{\rm B}T}{2e^2}\int\mathrm{d}^3\br\;\e(\br)\left\{-\left[\nabla\phi\ce(\br)\right]^2+\left[\nabla\psi(\br)\right]^2\right\}\nonumber\\
&&+\int\mathrm{d}^3\br\sigma(\br)\left[\phi\ce(\br)-i\psi(\br)\right]\nonumber\\
&&+i\frac{k_{\rm B}T}{e^2}\int\mathrm{d}^3\br\;\e(\br)\left[\nabla\phi\ce(\br)\right]\cdot\left[\nabla\psi(\br)\right].
\eea
At this point, we recall the Gauss' law 
\be
\label{h2}
\int_V\mathrm{d}^3\br\;\nabla\cdot\left(f\bA\right)=\int_{S(V)}\mathrm{d}^2\bs\cdot\bA
\ee
for the general scalar function $f$ and the vector field $\bA$, where $S(V)$ stands for the surface of the system boundary enclosing the system volume $V$. Plugging the identity $\nabla\cdot\left(f\bA\right)=f\nabla\cdot\bA+\bA\cdot\nabla f$ into Eq.~(\ref{h2}), one obtains
\be
\label{h3}
\int_V\mathrm{d}^3\br\bA\cdot\nabla f=\int_{S(V)}\mathrm{d}^2\bs\cdot\left(f\bA\right)-\int\mathrm{d}^3\br f\nabla\cdot\bA.
\ee
Setting in Eq.~(\ref{h3}) $\bA=\epsilon(\br)\nabla\phi\ce(\br)$ and $f=\psi(\br)$, the integral in the third line of Eq.~(\ref{h1}) becomes
\bea
\label{h4}
&&\int\mathrm{d}^3\br\;\e(\br)\left[\nabla\phi\ce(\br)\right]\cdot\left[\nabla\psi(\br)\right]\\
&&=\int\mathrm{d}^2\bs\cdot\left[\psi\e\nabla\phi\ce\right]_S-\int\mathrm{d}^3\br\;\psi(\br)\nabla\cdot\left[\epsilon\nabla\phi\ce\right]_{\br},\nonumber
\eea
Due to the global electroneutrality condition, the surface field on the r.h.s. of Eq.~(\ref{h4}) vanishes, i.e. $\left.\nabla\phi\ce(\br)\right|_{S}=0$. Thus, Eq.~(\ref{h1}) takes the form
\bea
\label{h5}
\beta H_0[\psi]&=&\frac{k_{\rm B}T}{2e^2}\int\mathrm{d}^3\br\;\e(\br)\left\{-\left[\nabla\phi\ce(\br)\right]^2+\left[\nabla\psi(\br)\right]^2\right\}\nonumber\\
&&+\int\mathrm{d}^3\br\sigma(\br)\phi\ce(\br)\\
&&-i\int\mathrm{d}^3\br\psi(\br)\left\{\frac{k_{\rm B}T}{e^2}\nabla\cdot\e(\br)\nabla\phi\ce(\br)+\sigma(\br)\right\}.\nonumber
\eea

We now note that owing to the Poisson identity~(\ref{sp2}), the third line of Eq.~(\ref{h5}) vanishes. Moreover, the use of Eq.~(\ref{h3}) with $\bA=\epsilon(\br)\nabla\phi\ce(\br)$ and $f=\phi\ce(\br)$ yields
\bea
\label{h6}
&&\int\mathrm{d}^3\br\;\e(\br)\left[\nabla\phi\ce(\br)\right]^2\\
&&=\int\mathrm{d}^2\bs\cdot\left[\phi\ce\e\nabla\phi\ce\right]_S-\int\mathrm{d}^3\br\;\phi\ce(\br)\nabla\cdot\left[\epsilon(\br)\nabla\phi\ce(\br)\right].\nonumber
\eea
Inserting Eq.~(\ref{h6}) into Eq.~(\ref{h5}), omitting the vanishing surface term, and using again the Poisson equation~(\ref{sp2}), the SC-level Hamiltonian finally simplifies to
\be
\label{h7}
\beta H_0[\psi]=\frac{k_{\rm B}T}{2e^2}\int\mathrm{d}^3\br\;\e(\br)\left[\nabla\psi(\br)\right]^2+\frac{1}{2}\int\mathrm{d}^3\br\sigma(\br)\phi\ce(\br).
\ee

\section{Calculating the integrals of the spherical solute charge in Eq.~(\ref{sc11})}
\label{X}

Here, we explain the evaluation of the integrals in Eq.~(\ref{sc11}). To this aim, this equation will be recast in an analytically manageable form. First, we express Eq.~(\ref{sc11}) as
\bea
\label{a21}
\hspace{-2mm}G(d)&=&\frac{1}{\sqrt\pi\alpha c_i}\int_0^d\frac{\mathrm{d}z}{d}\int_{b_0}^{d-b_0}\mathrm{d}z'\int_{b_0}^{b_+(z')}\mathrm{d}b\;e^{-\frac{b^2}{\alpha^2}}\\
\hspace{-2mm}&&\hspace{1.5cm}\times\int\mathrm{d}^2\br'_\pa n_{\rm s}\left(\sqrt{\bu_\pa^2+(z'-z)^2}-b\right),\nonumber
\eea
with $\bu_\pa=\br'_\pa-\br_\pa$. Then, in order to exploit the translational symmetry along the membrane walls, we introduce the change of variable $\br'_\pa\to\bu_\pa$. Eq.~(\ref{a21}) becomes
\bea
\label{a22}
G(d)&=&\frac{2\sqrt{\pi}}{\alpha c_i}\int_0^d\frac{\mathrm{d}z}{d}\int_{b_0}^{d-b_0}\mathrm{d}z'\int_{b_0}^{b_+(z')}\mathrm{d}b\;e^{-\frac{b^2}{\alpha^2}}\\
&&\hspace{.7cm}\times\int_0^\infty\mathrm{d}u_\pa u_\pa\;n_{\rm s}\left(\sqrt{u_\pa^2+(z'-z)^2}-b\right).\nonumber
\eea
At this point, we carry out the second change of integration variable $u_\pa\to t=\sqrt{u_\pa^2+(z'-z)^2}$. Eq.~(\ref{a22}) simplifies to
\bea
\label{a23}
G(d)&=&\frac{2\sqrt{\pi}}{\alpha c_i}\int_0^d\frac{\mathrm{d}z}{d}\int_{b_0}^{d-b_0}\mathrm{d}z'\int_{b_0}^{b_+(z')}\mathrm{d}b\;e^{-\frac{b^2}{\alpha^2}}\\
&&\hspace{3.5cm}\times\int_{|z'-z|}^\infty\mathrm{d}tt\;n_{\rm s}\left(t-b\right).\nonumber
\eea
Next, we plug into Eq.~(\ref{a23}) the charge structure factor~(\ref{sc12}), and carry out the integral over the variable $t$. This yields
\bea
\label{a24}
G(d)&=&\frac{1}{2\sqrt{\pi}\alpha c_i}\int_0^d\frac{\mathrm{d}z}{d}\int_{b_0}^{d-b_0}\mathrm{d}z'\theta\left[b_+(z')-b_-(z',z)\right]\nonumber\\
&&\hspace{1.5cm}\times\int_{b_-(z',z)}^{b_+(z')}\mathrm{d}bb^{-1}\;e^{-\frac{b^2}{\alpha^2}},
\eea
where we used the function~(\ref{au1}) and introduced the additional auxiliary function $b_-(z',z)={\rm max\left(b_0,|z'-z|\right)}$.

In order to evaluate Eq.~(\ref{a24}), one has to eliminate the conditional functions. This can be achieved by switching the order of the integrals over the variables $z$ and $z'$, and expressing Eq.~(\ref{a24}) in the following piecewise form
\be
\label{a26}
G(d)=\frac{2\sqrt{\pi}}{\alpha c_id}\left[J_1+J_2+J_3+J_4\right],
\ee
where the integrals have been defined as
\begin{widetext}
\bea\label{a27}
J_1&=&\int_{b_0}^{d/2}\mathrm{d}z'\left\{\int_0^{z'-b_0}\mathrm{d}z\int_{z'-z}^{z'}\mathrm{d}b+\int_{z'-b_0}^{z'}\mathrm{d}z\int_{b_0}^{z'}\mathrm{d}b\right\}b^{-1}e^{-\frac{b^2}{\alpha^2}};\\
\label{a28}
\hspace{-1cm}J_2&=&\int_{b_0}^{d/2}\mathrm{d}z'\left\{\int_{z'}^{z'+b_0}\mathrm{d}z\int_{b_0}^{z'}\mathrm{d}b+\int_{z'+b_0}^d\mathrm{d}z\;\theta(2z'-z)\int_{z-z'}^{z'}\mathrm{d}b\right\}b^{-1}e^{-\frac{b^2}{\alpha^2}};\\
\label{a29}
\hspace{-1cm}J_3&=&\int_{d/2}^{d-b_0}\mathrm{d}z'\left\{\int_0^{z'-b_0}\mathrm{d}z\;\theta(d+z-2z')\int_{z'-z}^{d-z'}\mathrm{d}b+\int_{z'-b_0}^{z'}\mathrm{d}z\int_{b_0}^{d-z'}\mathrm{d}b\right\}b^{-1}e^{-\frac{b^2}{\alpha^2}};\\
\label{a30}
\hspace{-1cm}J_4&=&\int_{d/2}^{d-b_0}\mathrm{d}z'\left\{\int_{z'}^{z'+b_0}\mathrm{d}z\int_{b_0}^{d-z'}\mathrm{d}b+\int_{z'+b_0}^d\mathrm{d}z\int_{z-z'}^{d-z'}\mathrm{d}b\right\}b^{-1}e^{-\frac{b^2}{\alpha^2}}.
\eea

Eq.~(\ref{a26}) is now in an analytically computable form. The incomplete gaussian integrals over the radius $b$ can be evaluated in terms of the error function~\cite{math}. After long but straightforward algebra, one finds $J_1=J_2=J_3=J_4$ and 
\be
\label{a31}
G(d)=\frac{1}{2c_i}\left\{{\rm erf}\left(\frac{d}{2\alpha}\right)-{\rm erf}\left(\frac{b_0}{\alpha}\right)-\frac{2\alpha}{\sqrt\pi d}\left[e^{-\frac{b_0^2}{\alpha^2}}-e^{-\frac{d^2}{4\alpha^2}}\right]\right\}.
\ee
\end{widetext}

\section{Terminal charge potentials~(\ref{g5}) and~(\ref{g8})}
\label{ter}

In this appendix, we derive the explicit form of the dielectric potentials~(\ref{g5}) and~(\ref{g8}) experienced by the solute charges. The calculation of these potentials requires the knowledge of the Green's function satisfying Eq.~(\ref{greq}). In Ref.~{\cite{netzvdw}, this equation has been solved in Fourier space, and the renormalized Green's function~(\ref{g6}) has been obtained in the form
\begin{widetext}
\be\label{g17II}
\delta v\ce(\br_\pa-\br'_\pa,z,z')=\ell_{\rm B}\Delta\int_0^\infty\mathrm{d}k\frac{ J_0(k|\br_\pa-\br'_\pa|)}{1-\Delta^2e^{-2kd}}\left\{e^{-k(z+z')}+e^{-k(2d-z-z')}+2\Delta e^{-2kd}\cosh\left[k(z-z')\right]\right\}.
\ee
In Eq.~(\ref{g17II}), $\Delta=(\e_{\rm w}-\e_{\rm m})/(\e_{\rm w}+\e_{\rm m})$ is the dielectric contrast parameter, and $J_0(x)$ stands for the Bessel function of the first kind~\cite{math}. Plugging now Eq.~(\ref{g17II}) into Eqs.~(\ref{g5}) and~(\ref{g8}), the solute self-energies follow for $j=\{p,e\}$ as
\be\label{g18II}
U_j(\ba,z)=\frac{\ell_{\rm B}}{2}\int_0^\infty\mathrm{d}k\frac{\Delta}{1-\Delta^2e^{-2kd}}\tu_j(\ba,z),
\ee
with the Fourier coefficients 
\bea\label{g19II}
\tu_p(\ba,z)&=&p_+^2\left[e^{-2kz}+e^{-2k(d-z)}+2\Delta e^{-2kd}\right]+e_+^2\left[e^{-2k(z+a_z)}+e^{-2k(d-z-a_z)}+2\Delta e^{-2kd}\right]\\&&+2p_+e_+J_0(ka_\pa)\left[e^{-k(2z+a_z)}+e^{-k(2d-2z-a_z)}+2\Delta e^{-2kd}\cosh(ka_z)\right];\nonumber\\
\label{g20II}
\tu_e(\ba,z)&=&p_+^2\left[e^{-2k(z+a_z)}+e^{-2k(d-z-a_z)}+2\Delta e^{-2kd}\right]+e_+^2\left[e^{-2kz}+e^{-2k(d-z)}+2\Delta e^{-2kd}\right]\\
&&+2p_+e_+J_0(ka_\pa)\left[e^{-k(2z+a_z)}+e^{-k(2d-2z-a_z)}+2\Delta e^{-2kd}\cosh(ka_z)\right],\nonumber
\eea
and $a_\pa=\sqrt{a^2-a_z^2}$. Thus, the derivative of the potential in the polarization component~(\ref{g15}) of the pressure reads
\be\label{g21}
\frac{\partial U_p(\ba,z)}{\partial d}=-\ell_{\rm B}\int_0^\infty\mathrm{d}k\frac{\Delta ke^{-2kd}}{\left(1-\Delta^2e^{-2kd}\right)^2}\tilde{T}(\ba,z),
\ee
where
\bea
\label{g22}\tilde{T}(\ba,z)&=&p_+^2\left[1+\Delta e^{-2kz}\right]^2e^{2kz}+e_+^2\left[1+\Delta e^{-2k(z+a_z)}\right]^2e^{2k(z+a_z)}\nonumber\\
&&+2p_+e_+J_0(ka_\pa)\left[e^{k(2z+a_z)}+2\Delta\cosh(ka_z)+\Delta^2e^{-k(2z+a_z)}\right].\nonumber
\eea
Finally, in the point charge limit $a\to0$, the self-energies in Eq.~(\ref{g18II}) and their derivative simplify to 
\bea
 \lim_{a\to0}U_{j=\{p,e\}}(\ba,z)&=&U_0(z)=\frac{q_+^2\ell_{\rm B}}{2}\int_0^\infty\frac{\mathrm{d}k\Delta}{1-\Delta^2e^{-2kd}}
\left[e^{-2kz}+e^{-2k(d-z)}+2\Delta e^{-2kd}\right];\\
\frac{\partial U_0(z)}{\partial d}&=&-q_+^2\ell_{\rm B}\int_0^\infty\frac{\mathrm{d}kk\Delta}{\left(1-\Delta^2e^{-2kd}\right)^2}
\left(1+\Delta e^{-2kz}\right)^2e^{-2k(d-z)}.
\eea

\end{widetext}


\begin{thebibliography}{99}
\bibitem{rem0} In order to simplify the notation, the numerical prefactor in the actual variance $\sqrt2 \alpha_i$ of the fluctuating solute size $b$ has been removed via the rescaling $\sqrt2 \alpha_i\to\alpha_i$.
\bibitem{rem2} In Refs.~\cite{NetzPr1} and~\cite{pincus}, the pressure data for point charges and dumbbell-like ions have been respectively reported against the dimensionless distances $\tilde d=d/\mu$ and $\tilde d'=d/\mu'$, with the GC lengths $\mu=1/(2\pi q_+\ell_{\rm B}\sigma_{\rm m})$ and $\mu'=1/(2\pi q\ell_{\rm B}\sigma_{\rm m})$ defined in terms of the solute valency $q_+$ and the terminal charge valency $q=q_+/2$. In order to compare these separate sets of data in terms of the same dimensionless distance $\tilde{d}$, in Fig.~\ref{fig5}(c), we reported the data of Ref.~\cite{pincus} by converting the dimensionless separation distances $\tilde d'=d/\mu'$ and solute sizes $\tilde a'=a/\mu'$ via a rescaling with the GC length $\mu$ of Ref.~\cite{NetzPr1}.
\end{thebibliography}
\end{document}